\documentclass[epj]{svjour}

\usepackage{dsfont}
\usepackage{graphicx}
\usepackage{epstopdf}  % allows to use pdf AND ps figures; use at prompt:  pdflatex --shell-escape <foo.tex>

\begin{document}
\title{The electromagnetic Sigma-to-Lambda hyperon transition form factors at low energies}
\author{Carlos Granados\inst{1,2} \and Stefan Leupold\inst{1} \and Elisabetta Perotti\inst{1}% etc
}                     % Do not remove
\authorrunning{Granados et al.}
\titlerunning{Hyperon transition form factors}
\institute{Institutionen f\"or fysik och astronomi, Uppsala Universitet, Box
516, S-75120 Uppsala, Sweden \and  Jefferson Lab, 12000 Jefferson Ave., Newport News, VA 23606, United States}
%\institute{Insert the first address here \and the second here}
%
\date{31.1.2017}
\abstract{
Using dispersion theory the low-energy electromagnetic form factors for the transition of a Sigma to a Lambda hyperon 
are related to the pion vector form factor. The additionally required input, i.e.\ the two-pion--Sigma--Lambda amplitudes 
are determined from relativistic next-to-leading-order (NLO) baryon chiral perturbation theory 
including the baryons from the octet and optionally from the decuplet. 
Pion rescattering is again taken into account by dispersion theory. 
%The obtained results are compared to the direct calculations of chiral perturbation theory. 
It turns out that the inclusion of decuplet baryons is 
not an option but a necessity to obtain reasonable results. The electric transition form factor remains very small in the 
whole low-energy region. The magnetic transition form factor depends strongly on one not very well determined low-energy 
constant of the NLO Lagrangian. One obtains reasonable predictive power if this low-energy constant is determined from 
a measurement of the magnetic transition radius. Such a measurement can be performed 
at the future Facility for Antiproton and Ion Research (FAIR).
\PACS{
      {13.40.Gp}{Electromagnetic form factors}   \and
      {11.55.Fv}{Dispersion relations}   \and
      {13.75.Gx}{Pion-baryon interactions} \and
      {11.30.Rd}{Chiral symmetries} 
     } % end of PACS codes
} %end of abstract
\maketitle

\section{Introduction}
\label{sec:intro}

The quest to understand the structure of matter does not stop with identifying the building blocks of a composite object. 
One wants to understand quantitatively how the respective building blocks interact and how they are distributed inside of this 
composite object. Some possible ways to explore the intrinsic structure of an object are 
\begin{itemize}
\item[(a)] to excite it, 
\item[(b)] to scatter on it, 
\item[(c)] to replace some of its building blocks by other, similar ones.  
\end{itemize}
In atomic physics all these techniques 
produced key insights and cross-checks of our understanding, for instance by studying the hydrogen spectrum --- related to (a),
by Rutherford scattering --- related to (b), or by studying systems with different atomic nuclei but the same number of 
electrons or electronic versus muonic atoms --- related to (c). 

To explore the structure of the nucleon one proceeds along similar lines. Concerning the excitation spectrum an increasing 
number of nucleon resonances has been isolated over the past decades \cite{Agashe:2014kda}. The motivation of the present work,
however, derives more from an interplay of the approaches (b) and (c). A huge body of information has been obtained 
from electron-nucleon scattering \cite{Punjabi:2015bba} and related observables --- with the most recent clue of an apparent 
difference in the proton charge radius as extracted from electronic or muonic 
hydrogen, respectively \cite{Pohl:2010zza,Carlson:2015jba}. The central objects 
are the electromagnetic form factors and the corresponding low-energy quantities: electric charge, magnetic moment, electric 
and magnetic radii. We note in passing that the non-trivial magnetic moment of the proton provided one of the first hints 
on the intrinsic structure of the proton \cite{1933ZPhy...85....4F}. If one flips the spin 
of one of the quarks inside the nucleon, one obtains a Delta baryon.\footnote{One might interpret this spin flip in the sense
of an excitation (a) or a replacement (c), but in this case this classification is merely language, not content.} 
The quantities extracted from the scattering reactions electron-nucleon to electron-Delta are the Delta-to-nucleon transition 
form factors. Extrapolating to the photon point one obtains the helicity amplitudes \cite{Agashe:2014kda}. 
The transition form factors provide complementary information about the structure of the nucleon (and the Delta) 
and have also been studied in some detail \cite{Pascalutsa:2006up}. 

The lightest quarks, up and down, provide the con\-stituent-quark content of nucleon and Delta. 
Yet there is one more comparatively
light quark, the strange quark. In the spirit of approach (c) one can ask what changes about the nucleon (and/or the Delta) 
if one or several up or down quarks are replaced by strange quarks. Historically, the such obtained states, the hyperons, 
were instrumental in revealing the quarks as the building blocks of the nucleons and other hadrons \cite{GellMann:1964nj}.
This suggests that the intrinsic structures of hyperons and nucleons are intimately related. 
Obviously, hyperon electromagnetic form factors and transition form factors contain complementary information to the 
nucleon and Delta form factors. Their knowledge would provide crucial tests for our current picture of the nucleon structure and 
therefore deepen our understanding. Yet, the experimental information about hyperon form factors is rather limited. 
Essentially only the magnetic moments of the octet hyperons are known (and, of course, their charges) \cite{Agashe:2014kda}.
For the decuplet-octet transitions not even the helicity amplitudes have been determined. 

Of course, this present limitation in the knowledge about hyperon form factors is caused by the fact that octet hyperons are not 
stable, but decay on account of the weak interaction \cite{Agashe:2014kda}. Therefore hyperon-electron scattering is 
experimentally very difficult to realize. Yet, the crossing symmetry of relativistic quantum fields provides a new angle. 
While electron-baryon scattering probes the form factors in the space-like region, 
hyperon form factors are accessible in the time-like\footnote{Since there is some confusion in the literature we define these 
phrases explicitly; time-like/space-like means: modulus of energy larger/smaller than modulus of three-momentum.} 
region for high and low energies. For high energies one can study electron-positron scattering reactions to a hyperon and an 
antihyperon. In principle, ``direct'' form factors and transition form factors are accessible here. 
For low energies one can extract transition form factors from the Dalitz decays $Y \to Y' \, e^+ e^-$ where $Y$ and $Y'$ 
denote two distinct hyperons. 
Of course, it is a shortcoming that the space-like region of the form factors is not easily accessible for hyperons. However, 
to some extent there is a compensation for it. The weak decays of the hyperons are self-analyzing in the sense that the 
angular distributions of the decay products give access to the spin properties without explicit polarization. Thus one might 
get an easier access to the various form factors as compared to the nucleon and Delta-nucleon cases.

In the present and forthcoming works we will address electromagnetic form factors of hyperons at low energies from the theory 
side. The calculations will cover the whole space- and time-like low-energy region, but at present the experimental significance
resides in the time-like Dalitz-decay region. Such electromagnetic decays of hyperons could be studied with high statistics 
at the future Facility for Antiproton and Ion Research (FAIR) at Darmstadt, Germany. There, hyperons will be copiously 
produced in $\bar p \, p$ (PANDA \cite{Lutz:2009ff}) 
and $p \, p$ (HADES\footnote{P.\ Salabura, private communication; see also \cite{Lorenz:2016qyg} and references therein.}) 
collisions. 
In the present work we study the only form factors in the octet sector that are 
connected to a Dalitz decay, namely the electric and the magnetic transition form factor of the neutral $\Sigma^0$ hyperon 
to the $\Lambda$ hyperon. These transition form factors are accessible by high-precision measurements of 
the decay $\Sigma^0 \to \Lambda \, e^+ e^-$. 

The main part of the present work deals with the calculation of these transition 
form factors. However, some discussion about the experimental feasibility is appropriate: The transition form factors are 
functions of the invariant mass of the dilepton, i.e.\ of the $e^+ e^-$ system. To resolve the shape of a form factor requires 
some range of invariant masses. For the Dalitz decay $\Sigma^0 \to \Lambda \, e^+ e^-$ the upper limit of available invariant masses 
is given by $m_{\Sigma^0}-m_\Lambda \approx 77\,$MeV. This is not very large as compared to typical hadronic scales. 
Thus, to extract even the electric or magnetic transition radius --- the first non-trivial aspect of a form factor --- 
requires a high experimental precision. In addition, the extraction of these radii from decay data relies on a proper 
understanding of the electromagnetic part. The lowest-order QED part is easily worked out. However, if the impact of the hyperon 
transition form factors/radii is numerically small, then radiative QED corrections compete with the hadronic form-factor effects.
This interplay will be explored in \cite{husek-leupold}. In the present work we concentrate on the hadronic part, 
the calculation of the hyperon electromagnetic form factors for the transition $\Sigma^0$ to $\Lambda$. 

Chiral perturbation theory ($\chi$PT) provides a model-independent approach to 
low-energy QCD \cite{Weinberg:1978kz,Gasser:1983yg,Gasser:1984gg,Scherer:2002tk,Scherer:2012xha}. Beyond the 
pseudo-Goldstone bosons it is possible to include the baryon octet and maybe the 
decuplet \cite{Pascalutsa:2005nd,Pascalutsa:2006up,Ledwig:2014rfa}, but it is unclear how to treat 
other hadronic states in a systematic, model-independent way. In the interaction of hadrons with electromagnetism the vector
mesons turn out to be very prominent \cite{sakuraiVMD}. For the isovector case the $\rho$ meson influences the electromagnetic 
structure down to rather low energies. Experimentally the $\rho$ meson shows up as a resonance in the p-wave pion phase shift
and in the pion form factor. Both quantities are nowadays known to high 
precision \cite{Colangelo:2001df,GarciaMartin:2011cn,Hanhart:2012wi,Schneider:2012ez}. 
Therefore one might pursue the 
strategy to marry purely hadronic $\chi$PT with the experimentally known pion form factor. Dispersion 
theory allows to combine these ingredients. This is similar
in spirit to \cite{Stollenwerk:2011zz,Hanhart:2013vba,Niecknig:2012sj,Schneider:2012ez,Kang:2013jaa}.
Concerning nucleon form factors see also \cite{Frazer:1960zzb,Mergell:1995bf,Hoferichter:2016duk}. 
In purely hadronic $\chi$PT we will 
explore the options to consider explicitly the decuplet states as active degrees of freedom or to include them 
only indirectly via the low-energy constants of the next-to-leading order Lagrangian. 

In the present work these ideas are applied to the $\Sigma^0$-to-$\Lambda$ transition form factors. 
In contrast to elastic form factors the transition has the advantage that it is purely isovector. Therefore it provides a 
good first test case for our formalism. A direct calculation 
in relativistic three-flavor $\chi$PT has been performed in \cite{Kubis:2000aa}. Therefore we can check the
accuracy of the obtained results before extending it to other more involved cases. As next steps one could address in the future 
the transition of the decuplet $\Sigma$ $(J^P={\frac32}^+)$ to the $\Lambda$ hyperon and of the $\Delta$ to the nucleon 
(for the latter case, see also \cite{Pascalutsa:2006up}). 
% take out Delta for paper version!!
Inclusion of the isosinglet part of electromagnetism opens the way for all elastic and transition form factors of octet and 
decuplet hyperons. Of course, at least for the calculations with the decuplet hyperons as initial states --- as appropriate 
for the corresponding Dalitz decays --- one has to use a version of $\chi$PT that includes the decuplet states as active degrees
of freedom. But, as we will see, the results obtained in the present work suggest this anyway. 

The rest of the paper is structured as follows: In the next section the theoretical ingredients are described in detail. 
Section \ref{sec:res} provides the results. Thereafter a summary and an outlook are presented. Appendices are added to discuss 
technical aspects and cross-checks which would interrupt the main text too much.

\section{Ingredients}
\label{sec:ingr}

\subsection{Dispersive representations}
\label{sec:disrep}

To apply dispersion theory we formally study the reaction $\Sigma^0 \, \bar \Lambda \to \gamma^*$, saturate the 
intermediate states by a pion pair and in the end extend
the amplitude to the kinematical region $\Sigma^0 \to \Lambda \, \gamma^*$. 
Technically this is along the lines described, e.g., in \cite{Kang:2013jaa} based on 
\cite{Omnes:1958hv,Mandelstam:1960zz}. We expect that the saturation of the inelasticity by a pion pair provides a good 
approximation for the transition form factors at low energies. 

The form factors are defined in \cite{Kubis:2000aa}. For our case of interest this reads
\begin{eqnarray}
  \langle 0 \vert j^\mu \vert \Sigma^0 \bar\Lambda \rangle & = & e \, 
  \bar v_\Lambda \, \left(
    \left( \gamma^\mu + \frac{m_\Lambda-m_\Sigma}{q^2} \, q^\mu \right) \, F_1(q^2) \right. 
  \nonumber \\ && \phantom{mmmmm} \left. 
    - \frac{i \sigma^{\mu\nu} \, q_\nu}{m_\Lambda + m_\Sigma} \, F_2(q^2) 
  \right) \, u_\Sigma
  \label{eq:defFF}
\end{eqnarray}
with
\begin{eqnarray}
  G_E(q^2) & := & F_1(q^2) + \frac{q^2}{(m_\Sigma + m_\Lambda)^2} \, F_2(q^2) \,, 
  \nonumber \\ 
  G_M(q^2) & := & F_1(q^2) + F_2(q^2)  \,.
  \label{eq:defFFEM}
\end{eqnarray}
$q^2$ denotes the square of the invariant mass of the virtual photon. 
$G_{E/M}$ is called electric/magnetic transition form factor, $F_{1/2}$ is called Dirac/Pauli transition form factor.
%Note that in \cite{Kubis:2000aa} the electric transition form factor was not explicitly defined. 
%The definition (\ref{eq:defFFEM}) is in line with the $G_E$ as used in the calculations 
%of \cite{Kubis:2000aa}.\footnote{B.\ Kubis, private communication}
The transition form factors are chosen such that they fit to the direct form factors that are commonly introduced for the 
baryon octet \cite{Kubis:2000aa}. The appearance of $1/q^2$ in (\ref{eq:defFF}) in connection with $F_1$ enforces the vanishing
of $F_1$ and therefore of $G_E$ at the photon point, i.e.\ $G_E(0)=0$. 

To determine $G_M(0)=F_2(0)$ we use the experimental result for the decay $\Sigma^0 \to \Lambda \, \gamma$. 
It is governed by the matrix element 
%(up to sign and $i$) 
\begin{eqnarray}
  \label{eq:mrealgam}
  {\cal M} = \bar u_\Lambda \, \frac{e i \sigma_{\mu\nu} q^\nu}{m_\Lambda+m_\Sigma} \, \kappa \, u_\Sigma \, \varepsilon^\mu
\end{eqnarray}
with $\kappa = G_M(0)$ \cite{Kubis:2000aa}. The decay width is given by
\begin{eqnarray}
  \label{eq:Gamrealgam}
  \Gamma_{\Sigma^0 \to \Lambda \gamma} = \frac{e^2 \, \kappa^2 \, (m_\Sigma^2-m_\Lambda^2)^3}{8 \pi \, m_\Sigma^3 \, (m_\Lambda+m_\Sigma)^2}
\end{eqnarray}
which leads to $\kappa \approx 1.98$ in agreement with the particle-data-group (PDG) value \cite{Agashe:2014kda} 
\begin{eqnarray}
  \label{eq:mupdg}
  \mu := \kappa \, \frac{e}{m_\Lambda+m_\Sigma} 
  = \underbrace{\kappa \, \frac{2 m_p}{m_\Lambda+m_\Sigma}}_{\approx 1.61} \, \frac{e}{2 m_P}  \,.
\end{eqnarray}

For later use we introduce the electric and magnetic transition radii \cite{Kubis:2000aa}:
\begin{eqnarray}
  \label{eq:defradiusEME}
  \langle r^2_{E} \rangle := 6 \left. \frac{d G_E(q^2)}{dq^2} \right\vert_{q^2 = 0} 
\end{eqnarray}
and
\begin{eqnarray}
  \label{eq:defradiusM}
  \langle r^2_{M} \rangle := \frac{6}{G_M(0)} \left. \frac{d G_M(q^2)}{dq^2} \right\vert_{q^2 = 0} \,.
\end{eqnarray}

For the dispersive representation of the form factors utilizing the two-pion intermediate state 
one needs a par\-tial-wave decomposition \cite{Jacob:1959at} and an evaluation of the form factors and of 
the four-point amplitude $\Sigma^0\,\bar\Lambda\,\pi^+\,\pi^-$ for different helicity states. It is convenient to work in 
the center-of-mass frame, choose the $z$ axis along the direction of motion of the $\Sigma^0$ and choose the $z$-$x$ plane as the
reaction plane. The corresponding spinors are 
explicitly given, e.g., in \cite{pesschr}. So basically one needs to evaluate 
$\bar v_\Lambda(-p_z,\lambda) \, \Gamma \, u_\Sigma(p_z,\sigma)$ where $\Gamma$ is an arbitrary spinor matrix and 
$\sigma$ and $\lambda$ denote the helicities. %(not the spin!)
Because of parity invariance it is sufficient to evaluate this object 
for the two cases $\sigma=\lambda=+1/2$ and $\sigma=-\lambda=+1/2$. Concerning the form factors, for a given combination of 
helicities one obtains an amplitude $F(q^2,\sigma,\lambda)$ that is a 
superposition of the two form factors. In turn one can reconstruct the form factors from combinations of these amplitudes.

In the center-of-mass frame all components of the current in (\ref{eq:defFF}) vanish for $\sigma=\lambda=+1/2$ except for 
$\mu=3$. One obtains
\begin{eqnarray}
  && F(q^2,+1/2,+1/2) = \nonumber \\
%  && \bar v_\Lambda(-p_z,+1/2) \, \left( 
%    \gamma^3 \, F_1(q^2) - \frac{i \sigma^{30} \, \sqrt{q^2}}{m_\Lambda + m_\Sigma} \, F_2(q^2) 
%  \right) \, u_\Sigma(p_z,+1/2) \nonumber \\
  && =  \bar v_\Lambda(-p_z,+1/2) \, \gamma^3 \, u_\Sigma(p_z,+1/2) \; G_E(q^2) \,.
  \label{eq:FFE}
\end{eqnarray}
For $\sigma=-\lambda=+1/2$ all components vanish except for $\mu=1,2$ which are just related by a factor of $i$. One finds 
\begin{eqnarray}
  \label{eq:FFM}
  && F(q^2,+1/2,-1/2) = \nonumber \\
%  && \bar v_\Lambda(-p_z,-1/2) \, \left( 
%    \gamma^1 \, F_1(q^2) - \frac{i \sigma^{10} \, \sqrt{q^2}}{m_\Lambda + m_\Sigma} \, F_2(q^2) 
%  \right) \, u_\Sigma(p_z,+1/2) \nonumber \\
  && =  \bar v_\Lambda(-p_z,-1/2) \, \gamma^1 \, u_\Sigma(p_z,+1/2) \; G_M(q^2) \,.
\end{eqnarray}

It is convenient and avoids kinematical singularities if one divides out the respective spinor coefficient and formulates 
dispersion relations directly for the 
electric and magnetic form factor. However, one should first consider for which pair of the four quantities 
$G_E$, $G_M$, $F_1$ and $F_2$ one would like to set up a (low-energy) dispersive representation. 
Concerning a direct $\chi$PT calculation it has been proposed in \cite{Kubis:2000aa} to use the Dirac and Pauli form factor 
$F_1$ and $F_2$. From the point of view of our helicity decomposition the electric and magnetic form factor seem to be more 
direct. In principle, if one has an excellent input for all these quantities, it should not matter. In reality, however, the 
relations (\ref{eq:defFFEM}) mix different powers of $q^2$ which is an issue in a necessarily truncated low-energy expansion 
in powers of momenta. In the present work we will use the electric and magnetic form factor as a starting point. 
We have briefly explored the option to start with dispersive representations for the Dirac and Pauli form factor, but 
with the next-to-leading-order input of chiral perturbation theory the results were less convincing. Clearly this deserves 
more detailed studies in the future.

We will mainly use the subtracted dispersion relations (see also \cite{Schneider:2012ez})
\begin{eqnarray}
  &&  G_{M/E}(q^2) =  G_{M/E}(0)  
  \nonumber \\ && {}
  +  \frac{q^2}{12\pi} \, \int\limits_{4 m_\pi^2}^\infty \frac{ds}{\pi} \, 
  \frac{T_{M/E}(s) \, p_{\rm c.m.}^3(s) \, F^{V*}_\pi(s)}{s^{3/2} \, (s-q^2-i \epsilon)}  \,. \phantom{m}
  \label{eq:dispbasic}  
\end{eqnarray}
The subtraction constants that appear in (\ref{eq:dispbasic}) can be adjusted to match the form factors at the 
photon point, $G_E(0)=0$, $G_M(0)=\kappa$. In line with the names for the form factors we will denote 
the corresponding amplitudes $T_E$ and $T_M$ by electric and magnetic scattering amplitude, 
respectively. 

We might also examine an unsubtracted version
\begin{eqnarray}
  G_{M/E}(q^2) = 
  \frac{1}{12\pi} \, \int\limits_{4 m_\pi^2}^\infty \frac{ds}{\pi} \, 
  \frac{T_{M/E}(s) \, p_{\rm c.m.}^3(s) \, F^{V*}_\pi(s)}{s^{1/2} \, (s-q^2-i \epsilon)}  
  \label{eq:dispbasicunsubtr}  
\end{eqnarray}
and explore to which extent the pion loop plus pion rescattering saturates the magnetic moment of the transition,
\begin{eqnarray}
  \kappa \stackrel{?}{=} 
  \frac{1}{12\pi} \, \int\limits_{4 m_\pi^2}^\infty \frac{ds}{\pi} \, 
  \frac{T_{M}(s) \, p_{\rm c.m.}^3(s) \, F^{V*}_\pi(s)}{s^{3/2} }  \,,
  \label{eq:dispbasicunsubtrkappa}  
\end{eqnarray}
or to which extent the dispersively calculated ``charge'' vanishes:
\begin{eqnarray}
  0 \stackrel{?}{=} 
  \frac{1}{12\pi} \, \int\limits_{4 m_\pi^2}^\infty \frac{ds}{\pi} \, 
  \frac{T_{E}(s) \, p_{\rm c.m.}^3(s) \, F^{V*}_\pi(s)}{s^{3/2} }  \,.
  \label{eq:dispbasicunsubtrnull}  
\end{eqnarray}
In general we expect that the subtracted dispersion relations work much better than the unsubtracted ones. An exact dispersive 
representation for the form factors would include all possible inelasticities. In our framework we use only the two-pion 
inelasticity. Thus we neglect for instance the inelasticities caused by four pions, 
by a kaon-antikaon pair, by a baryon-antibaryon pair, \ldots. In practice these mesonic inelasticities start at 
$\sqrt{s} \approx$ 1 GeV and the baryonic ones at around 2 GeV; 
see also the corresponding discussion in \cite{Hoferichter:2016duk}. Thus, all these inelasticities except for the one caused 
by two pions are ``high-energy inelasticities''. If we limit ourselves
to low values of $q^2$, then the influence of these high-energy inelasticities is suppressed by powers of $1/s$. The more 
subtractions one uses in the dispersive representation, the higher the suppression of the unaccounted high-energy inelasticities.
Thus we have more trust in the subtracted dispersion relations (\ref{eq:dispbasic}) than in (\ref{eq:dispbasicunsubtr}). 
If we found in practice a semi-quantitative agreement for the unsubtracted dispersion relations (\ref{eq:dispbasicunsubtrkappa})
and (\ref{eq:dispbasicunsubtrnull}), then we would assume that the subtracted dispersion relations work well on a quantitative 
level. On the other hand, the subtracted dispersion relations are sufficient to deduce low-energy quantities like radii --- 
(\ref{eq:defradiusEME}), (\ref{eq:defradiusM}) --- and curvatures. The general
philosophy is that low-energy structures, i.e.\ variations in energy, are mainly caused by low-energy physics, 
the two-pion intermediate states. 

In the dispersive formulae the quantity $F^V_\pi$ denotes the pion form factor defined by
\begin{equation}
  \label{eq:pionFFdef}
  \langle 0 \vert j^\mu \vert \pi^+(p_+) \, \pi^-(p_-) \rangle = e \, (p_+^\mu -p_-^\mu) \, F^V_\pi((p_++p_-)^2)  \,. \phantom{m}
\end{equation}
$T_{E/M}$ is the reduced amplitude for the reaction 
$\Sigma^0\,\bar\Lambda \to \pi^+\,\pi^-$ projected on 
$J=1$, i.e.
\begin{eqnarray}
  && {\cal M}(s,\theta,+1/2,+1/2) = \nonumber \\ && 
  {}\bar v_\Lambda(-p_z,+1/2) \, \gamma^3 \, u_\Sigma(p_z,+1/2) \; p_{\rm c.m.} \, 
  T_E(s) \, \cos \theta \nonumber \\ && {}+ 
  \mbox{other partial waves, $J\neq 1$,} 
  \label{eq:projampM1}
\end{eqnarray}
and
\begin{eqnarray}
  && {\cal M}(s,\theta,+1/2,-1/2) =  \nonumber \\ && 
  {} \bar v_\Lambda(-p_z,-1/2) \, \gamma^1 \, u_\Sigma(p_z,+1/2) \; p_{\rm c.m.} \, 
  T_M(s) \, \sin \theta \nonumber \\ && {}+ 
  \mbox{other partial waves, $J\neq 1$.} 
  \label{eq:projampM2}
\end{eqnarray}
$p_{\rm c.m.}$ is the pion center-of-mass momentum. 
Inverting (\ref{eq:projampM1}) and (\ref{eq:projampM2}) yields \cite{Jacob:1959at}
\begin{eqnarray}
  && T_E(s) = \nonumber \\ &&
  \frac32 \, \int\limits_0^\pi d\theta \, \sin\theta \, 
  \frac{{\cal M}(s,\theta,+1/2,+1/2)}{\bar v_\Lambda(-p_z,+1/2) \, \gamma^3 \, u_\Sigma(p_z,+1/2) \; p_{\rm c.m.}} \, \cos\theta
  \nonumber \\ &&
  \label{eq:projE}  
\end{eqnarray}
and 
\begin{eqnarray}
  && T_M(s) = \nonumber \\ &&
  \frac34 \, \int\limits_0^\pi d\theta \, \sin\theta \, 
  \frac{{\cal M}(s,\theta,+1/2,-1/2)}{\bar v_\Lambda(-p_z,-1/2) \, \gamma^1 \, u_\Sigma(p_z,+1/2) \; p_{\rm c.m.}} \, \sin\theta \,.
  \nonumber \\ &&
  \label{eq:projM}  
\end{eqnarray}
In practice these formulae are used for the bare input, not for the full amplitudes that contain pion rescattering. 

\begin{figure}[h]
  \centering
  \begin{minipage}[c]{0.15\textwidth}
    \includegraphics[keepaspectratio,width=\textwidth]{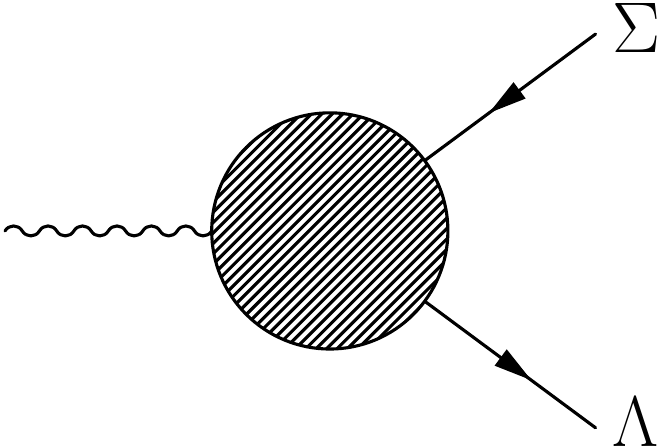}
  \end{minipage}  \hfill $\to$ \hfill 
  \begin{minipage}[c]{0.21\textwidth}  
    \includegraphics[keepaspectratio,width=\textwidth]{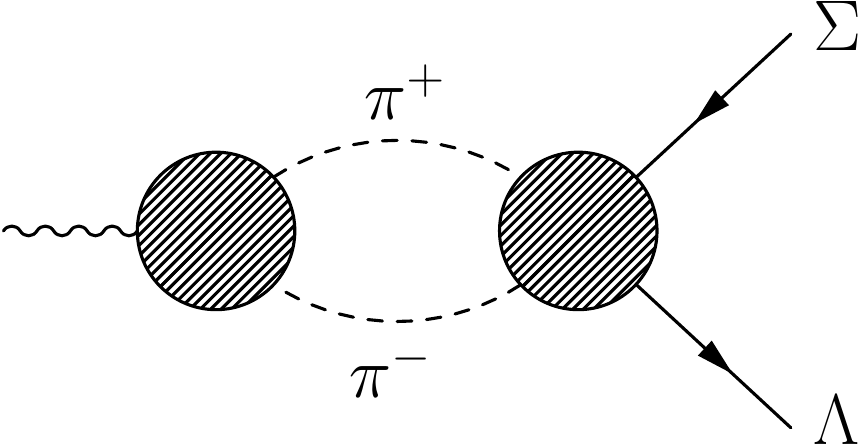}
  \end{minipage}  
  \caption{The transition form factors are obtained from their two-pion inelasticity.}
  \label{fig:SLTFFand2pi}
\end{figure}

Schematically the dispersion relation is depicted in figure \ref{fig:SLTFFand2pi}.

\begin{figure}[h]
  \centering
  \begin{minipage}[c]{0.15\textwidth}
    \includegraphics[keepaspectratio,width=\textwidth]{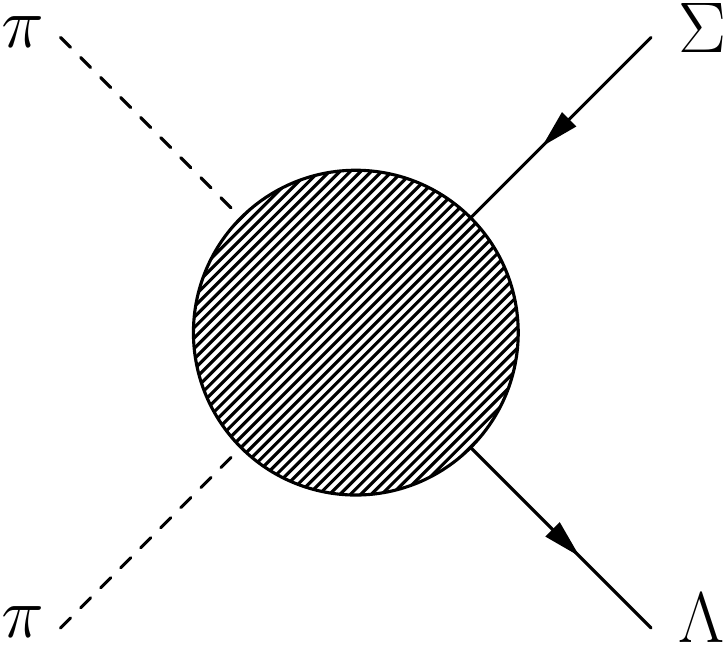}
  \end{minipage}  \hfill $\to$ \hfill 
  \begin{minipage}[c]{0.24\textwidth}  
    \includegraphics[keepaspectratio,width=\textwidth]{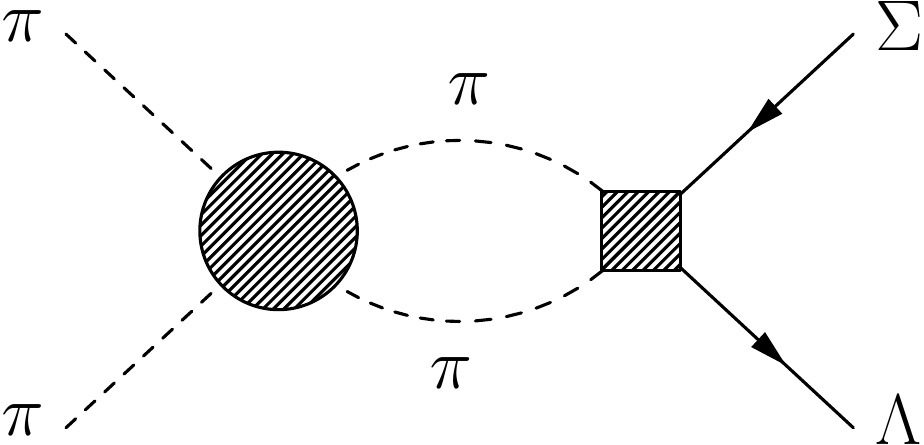}
  \end{minipage}  
  \caption{The scattering amplitude is obtained from the two-pion rescattering and a part (box) containing only left-hand cuts 
    and a polynomial.}
  \label{fig:SL2piand2pi}
\end{figure}

For the amplitude $T_{E/M}$ one should also consider 
pion rescattering encoded in the Omn\`es function
\begin{eqnarray}
  \Omega(s) = \exp\left\{ s \, \int\limits_{4m_\pi^2}^\infty \frac{ds'}{\pi} \, \frac{\delta(s')}{s' \, (s'-s-i \epsilon)} \right\}
  \approx F^V_\pi(s) 
  \label{eq:omnesele}  
\end{eqnarray}
where $\delta$ denotes the pion p-wave phase shift \cite{Colangelo:2001df,GarciaMartin:2011cn}. 
This is depicted in figure \ref{fig:SL2piand2pi}.
In practice we will follow the recipe of \cite{Hanhart:2012wi} and parametrize the phase shift such that it smoothly reaches 
$\pi$ at infinity. Contrary to \cite{Hanhart:2012wi} we do not include other inelasticities in the pion form factor, i.e.\ 
we do not distinguish between $\Omega$ and $F^V_\pi$. For our low-energy calculation this should not matter too much. Indeed 
we will see that other uncertainties are more severe. 

Along the lines of \cite{Mandelstam:1960zz,Kang:2013jaa} one needs an approximation for the 
``bare'' four-point amplitude $K$ of $\Sigma^0\,\bar\Lambda \to \pi^+\,\pi^-$ where pion rescattering
is ignored. In other words one needs the left-hand cuts of this amplitude. Ideally one would like to obtain this amplitude from 
(dispersion theory and) data from the crossed channel, i.e.\ from hyperon-pion scattering. Indeed, for the corresponding 
isovector part of the nucleon form factors such an analysis has been performed recently \cite{Hoferichter:2016duk} based 
on a dispersive Roy-Steiner analysis of pion-nucleon scattering \cite{Hoferichter:2015hva}. Since pions and hyperons are 
unstable, data on pion-hyperon scattering will not be available in the near future. For a coupled-channel analysis of 
pion-nucleon and kaon-nucleon scattering data with hyperons at least in the final states see \cite{Lutz:2001yb}. 
In lack of pion-hyperon scattering data we resort to the second best option and use 
in the following relativistic three-flavor 
$\chi$PT at next-to-leading order (NLO) to determine $K$. Strictly speaking the reaction amplitude for 
$\Sigma^0\,\bar\Lambda\to\pi^+\,\pi^-$ does not exist in baryon $\chi$PT, because 
there are no antibaryons in this framework. But the cross-channel amplitude $\Sigma^0\,\pi^+ \to \Lambda\,\pi^+$ does exist
and crossing symmetry and analytical continuation will provide the right answer. 

Given any input for $K$, the scattering amplitude $T$ is obtained by \cite{Kang:2013jaa} 
\begin{eqnarray}
  T(s) & = & K(s) + \Omega(s) \, P_{n-1}(s) \nonumber \\ && {}+ \Omega(s) \, s^n \, 
    \int\limits_{4m_\pi^2}^\infty \, \frac{ds'}{\pi} \, 
    \frac{\sin\delta(s') \, K(s')}{\vert\Omega(s')\vert \, (s'-s-i \epsilon) \, {s'}^n} \phantom{m}
  \label{eq:tmandel}
\end{eqnarray}
where $P_{n-1}$ denotes a polynomial of degree $n-1$. 

Note that any polynomial part of $K$ can be put into $P_{n-1}$ and need not be carried through the dispersion integral.
Thus one can split up the calculated Feynman amplitudes into a part ${\cal M}^{\rm pole}$ that contains the left-hand cuts --- 
in practice they will emerge from the pole terms of $u$- and $t$-channel exchange diagrams --- and a part ${\cal M}^{\rm contact}$ 
that is purely polynomial. Recalling the projection formulae (\ref{eq:projE}) and (\ref{eq:projM}) this leads to
\begin{eqnarray}
  && K_E(s) = \nonumber \\ && 
  \frac32 \, \int\limits_0^\pi d\theta \, \sin\theta \, 
  \frac{{\cal M}^{\rm pole}(s,\theta,+1/2,+1/2)}{\bar v_\Lambda(-p_z,+1/2) \, \gamma^3 \, u_\Sigma(p_z,+1/2) \; p_{\rm c.m.}} 
  \, \cos\theta  \nonumber \\ && 
  \label{eq:KprojE}  
\end{eqnarray}
and
\begin{eqnarray}
  && P_{n-1}^E(s) = \nonumber \\ && 
  \frac32 \, \int\limits_0^\pi d\theta \, \sin\theta \, 
  \frac{{\cal M}^{\rm contact}(s,\theta,+1/2,+1/2)}{\bar v_\Lambda(-p_z,+1/2) \, \gamma^3 \, u_\Sigma(p_z,+1/2) \; p_{\rm c.m.}} 
  \, \cos\theta  \nonumber \\ && 
  \label{eq:PprojE}  
\end{eqnarray}
and the equivalent formulae for the magnetic part. 

\begin{figure}[h]
  \centering
  \begin{minipage}[c]{0.16\textwidth}
    \includegraphics[keepaspectratio,width=\textwidth]{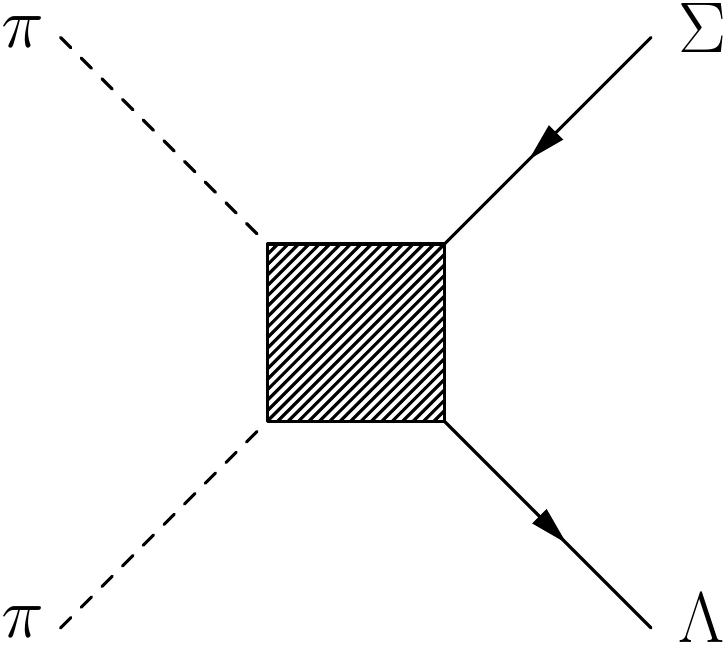}
  \end{minipage}  \hfill $\approx$ \hfill 
  \begin{minipage}[c]{0.12\textwidth}  
    \includegraphics[keepaspectratio,width=\textwidth]{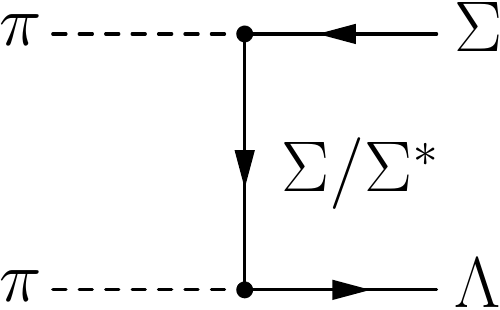}
  \end{minipage}   \hfill $+$ \hfill 
  \begin{minipage}[c]{0.07\textwidth}  
    \includegraphics[keepaspectratio,width=\textwidth]{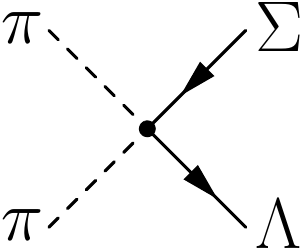}
  \end{minipage}  
  \caption{The ``bare'' input (box) is obtained from NLO $\chi$PT.}
  \label{fig:SL2pilhc}
\end{figure}

As already spelled out we will use three-flavor $\chi$PT to determine $K$ and the polynomial $P_{n-1}$.
Two versions are conceivable. One might or might not include the decuplet states explicitly. We will explore both options in 
the following. In any case we will restrict ourselves to NLO. 
As will be discussed below, leading order (LO) boils down to the
exchange diagrams $\pi^+\,\Sigma^0 \to \Sigma^+ \to \pi^+ \, \Lambda$ and (optionally) 
$\pi^+\,\Sigma^0 \to \Sigma^{*+} \to \pi^+ \, \Lambda$ 
($s$ and $u$ channel --- or, concerning $\Sigma^0\,\bar\Lambda \to \pi^+\,\pi^-$, $t$ and $u$ channel). 
Here $\Sigma^*$ denotes a decuplet state. The coupling constant of the latter can be adjusted to the 
measured decay widths $\Sigma^* \to \pi \, \Lambda$ or $\Sigma^* \to \pi \, \Sigma$; see further discussion below. 
NLO adds just contact terms (and provides the flavor splitting that leads to the physical masses of the states instead of 
one averaged mass per multiplet). If the decuplet states are not included explicitly, then the size of the NLO contact terms 
is modified such that the static version of the decuplet exchange is implicitly accounted for \cite{Meissner:1997hn}.  
Loops appear only at next-to-next-to leading order (NNLO). They would bring in additional left-hand cuts. 
Our approximation for the input is depicted in figure \ref{fig:SL2pilhc}.

In formula (\ref{eq:tmandel}) the pion loop starts to contribute when $\Omega(s)$ deviates from unity. 
This happens at order $s$. Therefore we cannot constrain the polynomial $P_{n-1}$ better than to a constant, if our input 
is restricted to tree level, i.e.\ NLO of $\chi$PT. In other words we have to use $n=1$ and drop all polynomial
terms of higher order. We will see below that the Born terms produce a polynomial of order 0, i.e.\ a constant. For 
the magnetic/electric part the NLO contact term produces a polynomial of 
order 0/1 --- see (\ref{eq:nlocontact1}), (\ref{eq:nlocontact2}) below. Thus one should keep the NLO 
contribution for the magnetic part, but not for the electric. 
All this is in line with the treatment of \cite{Kubis:2000aa} 
as described in detail in Kubis' Ph.D.\ thesis \cite{KubisPhD} in the following sense. In a direct $\chi$PT calculation of the 
form factors the NLO contact term contributes only to the Pauli form factor $F_2$ \cite{KubisPhD}. 
On account of (\ref{eq:defFFEM}) the impact of $F_2$ on $G_E$ relative to $G_M$ is suppressed for low $q^2$. 

As we will discuss below, the decuplet-exchange terms yield polynomials which depend on the spurious spin-1/2 contributions. 
For the electric part the ambiguity is of second chiral order which should be dropped anyway.
For the magnetic part there is a constant term which can be accounted for equally well by the NLO contact term. Thus the 
polynomial part formally emerging from the decuplet exchange can be entirely dropped for the magnetic contribution. 
To obtain the proper low-energy 
limit of $\chi$PT we should use 
\begin{eqnarray}
  P_0^E & = & P^E_{\rm Born} + P^E_{\rm res}   \,, \nonumber \\
  P_0^M & = & P^M_{\rm Born} + P^M_{\rm NLO\, \chi PT} - K_{\rm res,low}^M  \,,
  \label{eq:fixpolyEM}
\end{eqnarray}
where the label ``Born'' denotes the Sigma exchange and ``res'' the exchange of the decuplet resonance. 
The label ``NLO $\chi$PT'' refers to $\chi$PT without the decuplet. $K_{\rm res,low}^M$ denotes 
the low-energy limit of the resonance-pole contribution to the magnetic amplitude. A detailed analysis reveals that there 
are some subtleties with this low-energy limit due to the left-hand cut structure of the resonance-exchange contributions.
This is discussed in detail in appendix \ref{sec:app-cutres} based on the results (\ref{eq:resfeyn}) below.

Note that 
the relation for the electric polynomial implies 
\begin{equation}
  P^E_{\rm res} + K_{\rm res,low}^E = \mbox{higher order}
  \label{eq:conslel}
\end{equation}
to be consistent with the low-energy limit of $\chi$PT. We have checked that this is indeed the case.

Let us briefly discuss the convergence of the integrals in (\ref{eq:dispbasic}) and (\ref{eq:tmandel}): 
If the pole terms from the Born diagrams (octet exchange) are projected on $J=1$, they scale like $(\log s)/s$ 
for large $s$. The 
subtracted dispersion relation in (\ref{eq:tmandel}) converges very well. The high-energy behavior of the curly bracket is then
$\sim s^0$. Since the Omn\`es function behaves like $1/s$, 
the whole amplitude scales like $(\log s)/s$ at large $s$. This provides a 
very convergent integral in (\ref{eq:dispbasic}). 
The decuplet changes the picture to some extent: The pole terms diverge like $\log s$. Still this leads to convergent integrals.

On the other hand, the analytic structure of the scattering amplitudes $K_{M/E}$ changes where $p_z$ or $p_{\rm c.m.}$ have 
their zeros. This happens at $s_1:=(m_\Sigma-m_\Lambda)^2$, $s_2:=4m_\pi^2$ and $s_3:=(m_\Sigma+m_\Lambda)^2$. 
We are interested in $q^2 < s_1$ for the transition form factors (\ref{eq:dispbasic}) and we try to obtain a reasonable 
approximation for the scattering amplitudes $K_{M/E}(s)$ and $T_{M/E}(s)$ in the low-energy part of $s_2 \le s < s_3$.
Thus it does not make sense to evaluate the functions outside of $s_2 \le s < s_3$.

In practice we will terminate the integration range in (\ref{eq:dispbasic}) and in (\ref{eq:tmandel}) by a 
finite cutoff $\Lambda^2$ and check the sensitivity of our results to a variation in $\Lambda$. 
From the previous considerations it is clear that we should keep the cutoff $\Lambda$ below $\sqrt{s_3}=m_\Sigma+m_\Lambda$. 
We will vary $\Lambda$ between 1 and 2 GeV and study the impact of this change on the results.

\subsection{Lagrangians, parameters and input tree-level amplitudes}
\label{sec:lagretc}

The relevant interaction part of the LO chiral lagrangian \cite{Kubis:2000aa} including only the octet baryons 
as active degrees of freedom is given by
\begin{eqnarray}
  {\cal L}_8^{(1)} &=& i \langle \bar B \gamma_\mu D^\mu B \rangle
  + \frac{D}{2} \, \langle \bar B \, \gamma^\mu \, \gamma_5 \, \{u_\mu,B\} \rangle  \nonumber \\
  && {} + \frac{F}{2} \, \langle \bar B \, \gamma^\mu \, \gamma_5 \, [u_\mu,B] \rangle  
  \label{eq:baryonlagr8}
\end{eqnarray}
with the octet baryons collected in
\begin{eqnarray}
  \label{eq:baroct}
  B = \left(
    \begin{array}{ccc}
      \frac{1}{\sqrt{2}}\, \Sigma^0 +\frac{1}{\sqrt{6}}\, \Lambda 
      & \Sigma^+ & p \\
      \Sigma^- & -\frac{1}{\sqrt{2}}\,\Sigma^0+\frac{1}{\sqrt{6}}\, \Lambda
      & n \\
      \Xi^- & \Xi^0 
      & -\frac{2}{\sqrt{6}}\, \Lambda
    \end{array}   
  \right)  \,,
\end{eqnarray}
the Goldstone bosons encoded in
\begin{eqnarray}
  \Phi &=&  \left(
    \begin{array}{ccc}
      \pi^0 +\frac{1}{\sqrt{3}}\, \eta 
      & \sqrt{2}\, \pi^+ & \sqrt{2} \, K^+ \\
      \sqrt{2}\, \pi^- & -\pi^0+\frac{1}{\sqrt{3}}\, \eta
      & \sqrt{2} \, K^0 \\
      \sqrt{2}\, K^- & \sqrt{2} \, {\bar{K}}^0 
      & -\frac{2}{\sqrt{3}}\, \eta
    \end{array}   
  \right) 
  \,, \label{eq:gold1} \\
  u^2 & := & U := \exp(i\Phi/F_\pi) \,, \quad u_\mu := i \, u^\dagger \, (\nabla_\mu U) \, u^\dagger = u_\mu^\dagger \,, \phantom{mm}
  \label{eq:gold}
\end{eqnarray}
and $\langle \ldots \rangle$ denoting a flavor trace. The chirally covariant derivatives are defined by
\begin{eqnarray}
  \label{eq:devder}
  D^\mu B := \partial^\mu B + [\Gamma^\mu,B]
\end{eqnarray}
with
\begin{eqnarray}
  \Gamma_\mu &:= & \frac12 \, \left(
    u^\dagger \, \left( \partial_\mu - i (v_\mu + a_\mu) \right) \, u \right. \nonumber \\ && \left. {}+
    u \, \left( \partial_\mu - i (v_\mu - a_\mu) \right) \, u^\dagger
  \right) \,,
  \label{eq:defGammamu}
\end{eqnarray}
and
\begin{eqnarray}
  \label{eq:defnabla}
  \nabla^\mu U := \partial^\mu U - i (v^\mu + a^\mu) \, U + i U \, (v^\mu - a^\mu)
\end{eqnarray}
where $v$ and $a$ denote external sources.

If one includes also the decuplet states as active degrees of freedom in $\chi$PT, then the relevant interaction part of 
the LO chiral lagrangian reads \cite{Jenkins:1991es,Pascalutsa:2006up,Ledwig:2014rfa} 
%(note that we use the relativistic version, for the power counting this would only be a potential problem for loop corrections; 
%also note that \cite{Pascalutsa:2005nd} has an inconsistent normalization condition for the isospin 3/2 to 1/2 transition 
%matrices, in \cite{Pascalutsa:2006up} it is correct) 
\begin{eqnarray}
  {\cal L}_{8+10}^{(1)} &=& {\cal L}_{8}^{(1)}   \nonumber \\ && {}
  + \frac{1}{2\sqrt{2}} \, h_A \, \epsilon_{ade} \, g_{\mu\nu} \, 
  (\bar T^\mu_{abc} \, u^\nu_{bd} \, B_{ce} + \bar B_{ec} \, u^\nu_{db} \, T^\mu_{abc})  \nonumber \\ 
  \label{eq:baryonlagr}
\end{eqnarray}
where the decuplet is expressed by a totally symmetric flavor tensor $T_{abc}$ 
with \cite{Ledwig:2014rfa} 
\begin{eqnarray}
  && T_{111} = \Delta^{++} \,, \quad T_{112} = \frac{1}{\sqrt{3}} \, \Delta^+  \,, \nonumber \\
  && T_{122} = \frac{1}{\sqrt{3}} \, \Delta^0  \,,  T_{222} = \Delta^- \,, \nonumber \\
  && T_{113} = \frac{1}{\sqrt{3}} \, \Sigma^{*+}  \,, \quad T_{123} = \frac{1}{\sqrt{6}} \, \Sigma^{*0}  \,, \quad 
  T_{223} = \frac{1}{\sqrt{3}} \, \Sigma^{*-}  \,,  \nonumber \\
  && T_{133} = \frac{1}{\sqrt{3}} \, \Xi^{*0} \,, \quad T_{233} = \frac{1}{\sqrt{3}} \, \Xi^{*-} \,, \quad %&& \nonumber \\
  T_{333} = \Omega \,. 
  \label{eq:tensorT}
\end{eqnarray}
%Note that the particle assignment in (\ref{eq:gold1}) is in disagreement with the standard phase conventions for the isospin 
%Clebsch-Gordan coefficients, see also equation (5.16) in \cite{Gasser:1984gg}. 

The last term in (\ref{eq:baryonlagr}) provides the pion-hyperon three-point interactions. Of course, it is not unique how to 
write down this 
interaction term \cite{Meissner:1997hn,Pascalutsa:1999zz,Lutz:2001yb,Pascalutsa:2005nd,Pascalutsa:2006up,Ledwig:2014rfa}. 
In principle, all differences can be encoded in the contact interactions that show up in $\chi$PT at NLO; see below. 
In practice, it might happen that different versions of the LO three-point interaction terms once used with physical 
masses induce flavor-breaking effects that are not entirely accounted for by NLO contact terms. From a formal point of view 
such effects are NNLO, but in practice it might matter to some extent; see also the discussion in \cite{Lutz:2001yb}. 
In the present work we are not interested in a description of all hyperon form factors, but focus on the $\Sigma^0$-to-$\Lambda$ 
transition. If one does not use or insist on cross-relations between NLO parameters induced by three-flavor symmetry, then 
all differences between different versions of the three-point interactions can be moved to the contact interactions. Below 
we will explore explicitly two versions of the LO three-point interaction term to substantiate our statements. 

For the coupling constants we use $F_\pi=92.4\,$MeV, $D=0.80$, $F=0.46$ \cite{Kubis:2000aa} and $h_A$ determined from
the partial decay width $\Sigma^* \to \pi \, \Lambda$ or $\Sigma^* \to \pi \, \Sigma$. 
The partial width for the decay of a decuplet state with mass $M$ into an octet state with mass $m$ plus a pion 
and with a coefficient $c$ in the lagrangian of type (\ref{eq:baryonlagr}) is given by
\begin{eqnarray}
  \label{eq:widthDelta}
  \Gamma = \frac{c^2}{12\pi} \, {\tilde p_{\rm c.m.}}^3 \, \frac{E_B+m}{M}
\end{eqnarray}
where $E_B=\sqrt{m^2+{\tilde p_{\rm c.m.}}^2}$ ($\tilde p_{\rm c.m.}$) is the energy (momentum) of the outgoing baryon in the 
rest frame of the decaying resonance. For the decays of interest one finds from the explicit interaction lagrangian 
(\ref{eq:baryonlagr}): $c_{\Sigma^*\Lambda\pi}=h_A/(2 \sqrt{2} F_\pi)$, and $c_{\Sigma^*\Sigma\pi}=h_A/(2 \sqrt{6} F_\pi)$. 
(Note that there are always two decay branches possible for each decay $\Sigma^*\to\Sigma\pi$.) Matching to the experimental 
results yields $h_A=2.4$ from $\Sigma^*\to\Lambda\pi$ and $h_A$ ranging between 
2.2 and 2.3 from $\Sigma^*\to\Sigma\pi$ --- here the mass differences between isospin partners matter!
For the numerical calculations we will explore the range 
\begin{equation}
  \label{eq:harange}
  h_A = 2.3 \pm 0.1  \,.
\end{equation}

We note in passing that one obtains a somewhat larger value for $h_A$ from the partial decay width $\Delta \to N\pi$.
Here $c_{\Delta N\pi}=h_A/(2 F_\pi)$ and $h_A=2.88$. Finally one might look at the large-$N_c$ prediction 
(see, e.g., \cite{Pascalutsa:2005nd} and references therein --- $N_c$ denotes the number of colors): 
$h_A=3 g_A/\sqrt{2}=2.67$  with $g_A=F+D =1.26$. 
In the following we will use $h_A$ for the vertices $\Sigma^* \Lambda \pi$ and $\Sigma^* \Sigma \pi$. Therefore we regard 
the determination from the $\Sigma^*$ decays as the most reasonable ones for our purposes. The difference to the determination
from the $\Delta$ decay points towards flavor breaking effects for this coupling which shows up at NNLO in the chiral counting.

According to \cite{Oller:2006yh} a complete and minimal NLO Lagrangian for the baryon-octet sector is given by 
\begin{eqnarray}
  {\cal L}_{8}^{(2)} &=& b_D \langle\bar{B} \{ \chi_+,B\}\rangle +
  b_F \langle\bar{B}[ \chi_+,B ]\rangle +
  b_0 \langle\bar{B} B\rangle\langle\chi_+\rangle \nonumber \\
  && {}+
  b_1 \langle\bar{B} [ u^\mu,[u_\mu,B]]\rangle +
  b_2 \langle\bar{B}\{ u^\mu,\{u_\mu,B\}\}\rangle \nonumber \\
  && {}+
  b_3 \langle\bar{B}\{ u^\mu,[u_\mu,B]\}\rangle +
  b_4 \langle\bar{B}B\rangle\langle u^\mu u_\mu\rangle \nonumber \\
  && {}+
  i b_5 \big(\langle\bar{B}[u^\mu,[u^\nu,\gamma_\mu
    {D}_\nu B]]\rangle  \nonumber \\
  && \phantom{mmmm} {}-
    \langle \bar{B}\overleftarrow{D}_\nu[u^\nu,[u^\mu,\gamma_\mu B]]\rangle\big)  \nonumber \\
  && {}+
  i b_6
  \big(\langle\bar{B}[u^\mu,\{u^\nu,\gamma_\mu{D}_\nu B\}]\rangle \nonumber \\
  && \phantom{mmmm} {}-
    \langle \bar{B}\overleftarrow{D}_\nu\{u^\nu,[u^\mu,\gamma_\mu B]\}\rangle\big)  \nonumber \\
  && {}+
  i b_7 \big(\langle\bar{B}\{u^\mu,\{u^\nu,\gamma_\mu{D}_\nu B\}\}\rangle \nonumber \\
  && \phantom{mmmm} {}-
    \langle \bar{B}\overleftarrow{D}_\nu \{u^\nu,\{u^\mu,\gamma_\mu B\}\}\rangle\big)  \nonumber \\
  && {}+
  i b_8\big(\langle\bar{B}\gamma_\mu{D}_\nu B\rangle-
    \langle \bar{B}\overleftarrow{D}_\nu\gamma_\mu B\rangle\big)\langle u^\mu u^\nu\rangle \nonumber \\
  && {}+
  \frac{i}{2}\, b_9 \, \langle\bar{B} u^\mu\rangle \langle u^\nu \sigma_{\mu\nu} B\rangle   \nonumber \\
  && {}+
  \frac{i}{2} \, b_{10} \, \langle\bar{B}\{[u^\mu,u^\nu],\sigma_{\mu\nu} B\}\rangle \nonumber \\
  && {}+
  \frac{i}{2} \, b_{11} \, \langle\bar{B}[[u^\mu,u^\nu],\sigma_{\mu\nu} B]\rangle    \nonumber \\
  && {}+
  d_4 \langle\bar{B}\{ f_+^{\mu\nu},\sigma_{\mu\nu} B\}\rangle +
  d_5 \langle\bar{B}[ f_+^{\mu\nu},\sigma_{\mu\nu} B]\rangle  \,.
  \label{eq:NLO1}
\end{eqnarray}
with $\chi_\pm = u^\dagger \chi u^\dagger \pm u \chi^\dagger u$ and $\chi = 2 B_0 \, (s+ip)$ 
obtained from the scalar source $s$ and the pseudoscalar source $p$. The low-energy constant $B_0$ is essentially the ratio of
the light-quark condensate to the square of the pion-decay constant. 

We note in passing that Frink and Mei\ss ner \cite{Frink:2006hx} agree with \cite{Oller:2006yh} at the NLO level displayed 
in (\ref{eq:NLO1}), though not at NNLO.
To be in line with the conventions of \cite{Kubis:2000aa} we have relabeled some of the coupling constants 
of \cite{Oller:2006yh}: $d_1 \to b_{10}/2$, $d_2 \to b_{11}/2$, $d_3 \to b_9/2$. 
The terms $\sim b_{D/F}$ provide the mass splitting for the octet states. 
Concerning the interaction terms for  
$\bar\Lambda \Sigma^0 \pi^+\pi^-$ only $b_D$, $b_3$, $b_6$, and $b_{10}$ contribute. A more detailed investigation reveals that 
the $b_6$ term is not of NLO in this channel. Concerning the scattering of baryon-antibaryon to 
two pions the $b_D$, $b_3$ terms do not contribute to the p-wave. Thus for our p-wave amplitudes we will only need 
a value for $b_{10}$. If we do not include the decuplet states as explicit degrees of freedom, we can take the value of 
$b_{10}$ from the corresponding works on $\chi$PT. In \cite{Meissner:1997hn} a value of $b_{10} \approx 0.95\,$GeV$^{-1}$ has 
been given. In \cite{Kubis:2000aa} a somewhat larger value is used, $b_{10} \approx 1.24 \,$GeV$^{-1}$. 
In our calculations we will explore the range 
\begin{equation}
  \label{eq:b10range}
  b_{10} = (1.1 \pm 0.25)\, {\rm GeV}^{-1} \,.
\end{equation}
In practice this is all we need to provide input for (\ref{eq:fixpolyEM}). 

To illuminate the meaning and input for the contact interactions we add the following discussion.
Unfortunately the value for $b_{10}$ is not entirely based on experimental
input. Instead a resonance saturation assumption enters the estimate for $b_{10}$ \cite{Meissner:1997hn}. 
In this framework a significant part of the value for $b_{10}$ comes from the contribution of the decuplet exchange. 
Thus if the decuplet baryons are included as active degrees of freedom the low-energy constants in the NLO lagrangian must 
be readjusted. We denote the NLO low-energy constants of octet+decuplet $\chi$PT by $\tilde b_{\dots}$ instead of $b_{\dots}$. 
Consequently the relevant part of the NLO Lagrangian for octet+decuplet $\chi$PT is given by
\begin{eqnarray}
  {\cal L}_{8+10}^{(2)} &=& \left. {\cal L}_{8}^{(2)} \right\vert_{b_{\dots} \to \tilde b_{\dots}} + \mbox{mass splitting for decuplet.}
  \phantom{mm}
  \label{eq:NLO110}
\end{eqnarray}
Note that this is not the complete NLO lagrangian of octet+decuplet $\chi$PT, only the part relevant for our purposes.

As already stressed, the only NLO low-energy constant that really matters for our calculations is $b_{10}$ or $\tilde b_{10}$, 
respectively. To relate these two quantities in the most reasonable way in view of the $\Sigma \Lambda \pi^+ \pi^-$ amplitude 
we have to determine the low-energy and/or chiral-limit contribution to this amplitude from the decuplet exchange 
(see further discussion below). If we denote 
this contribution by $b_{10}^{\rm res}$ we have to choose $\tilde b_{10}$ such that the sum produces the result of pure 
baryon-octet $\chi$PT:
\begin{eqnarray}
  \label{eq:b10det}
  \tilde b_{10} + b_{10}^{\rm res} = b_{10}  \,.
\end{eqnarray}
On the other hand, if we are not interested in an explicit value for $\tilde b_{10}$ we can just use (\ref{eq:fixpolyEM}). 

Alternatively to the resonance saturation of \cite{Meissner:1997hn} one might utilize input from \cite{Lutz:2001yb}. 
There, scattering data on pion-nucleon 
and kaon-nucleon have been described by a chiral coupled-channel Bethe-Salpeter approach. In this framework the 
contact interactions 
have been determined from large-$N_c$ constraints and fits to the scattering data. We have checked explicitly that these 
contact interactions can be translated to a $b_{10}$ parameter that is in the range given in (\ref{eq:b10range}). Thus in practice
we use (\ref{eq:fixpolyEM}) together with (\ref{eq:b10range}). 

For the tree-level calculation of the four-point amplitude $\pi^+\,\pi^-\,\Sigma^0\,\bar\Lambda$ there can be exchange 
contributions from the three-point vertices $\sim D, F, h_A$ and contact interactions from the NLO terms. 
In addition, one might get a contact term of Weinberg-Tomozawa type \cite{Weinberg:1966kf,Tomozawa:1966jm} 
from the chiralized kinetic term, i.e.\ from tr$(\bar B \gamma^\mu [\Gamma_\mu,B])$. However, if one considers only pions, the 
non-trivial part of $\Gamma_\mu$ resides only in the first two rows and columns. There, however, the $\Lambda$ part of $B$ is 
proportional to the unit matrix. Therefore, the commutator $[\Gamma_\mu,B]$ vanishes and there is no Weinberg-Tomozawa term for 
the four-point amplitude we are interested in. (Considering $\bar\Lambda$ for $\bar B$ and
$\Sigma^0$ for $B$ does not change the argument, because one can shift the commutator to $\bar B$ within the flavor trace.)

The Feynman amplitude for the Born terms (exchange of octet $\Sigma^\pm$) is given by
\begin{eqnarray}
  && {\cal M}_{\rm Born}  = \frac{D \, F}{\sqrt{3} F_\pi^2} \, \nonumber \\ && \times \Bigg(
    \bar v_\Lambda u_\Sigma \, m_\Sigma \, (m_\Sigma^2-m_\Lambda^2) \, 
    \left( \frac{1}{t-m_\Sigma^2}-\frac{1}{u-m_\Sigma^2} \right)  \nonumber \\ && \phantom{m} {}
    + \bar v_\Lambda \gamma^\mu k_\mu u_\Sigma \, m_\Sigma \, (m_\Sigma+m_\Lambda) \, 
    \left( \frac{1}{t-m_\Sigma^2}+\frac{1}{u-m_\Sigma^2} \right)\nonumber \\ &&  \phantom{m} {}
    + \bar v_\Lambda \gamma^\mu k_\mu u_\Sigma 
  \Bigg)
  \label{eq:bornfeyn}
\end{eqnarray}
where $k = p_+-p_-$ denotes the difference of the pion momenta and $t=(p_\Sigma-p_+)^2$. In the center-of-mass system we have 
$k=(0,2 \, p_{\rm c.m.}\sin\theta,0,2 \, p_{\rm c.m.}\cos\theta)$. In line with (\ref{eq:projampM1}), (\ref{eq:projampM2}) we 
introduce reduced amplitudes. One obtains for the electric case, $\sigma=\lambda=+1/2$:
\begin{eqnarray}
  && \frac{{\cal M}_{\rm Born}}{\bar v_\Lambda(-p_z,+1/2) \, \gamma^3 \, u_\Sigma(p_z,+1/2) \; p_{\rm c.m.}} =  
  \frac{D \, F}{\sqrt{3} F_\pi^2} \, \nonumber \\ && \times \Bigg(
    - \frac{s-(m_\Sigma+m_\Lambda)^2}{2  \, p_{\rm c.m.} \, p_z} \, m_\Sigma \, (m_\Sigma-m_\Lambda) \, 
    \nonumber \\ && \phantom{mmmm} {} \times
    \left( \frac{1}{t-m_\Sigma^2}-\frac{1}{u-m_\Sigma^2} \right)  \nonumber \\ && \phantom{mm} {}
    - 2 \cos\theta \, m_\Sigma \, (m_\Sigma+m_\Lambda) \, 
    \left( \frac{1}{t-m_\Sigma^2}+\frac{1}{u-m_\Sigma^2} \right)\nonumber \\ &&  \phantom{mm} {}
    - 2 \cos\theta 
  \Bigg)   \,.
  \label{eq:bornfeynnorm1}
\end{eqnarray}
For the magnetic case, $\sigma=-\lambda=+1/2$, one finds:
\begin{eqnarray}
  && \frac{{\cal M}_{\rm Born}}{\bar v_\Lambda(-p_z,-1/2) \, \gamma^1 \, u_\Sigma(p_z,+1/2) \; p_{\rm c.m.}} =  
  \frac{D \, F}{\sqrt{3} F_\pi^2} \, \nonumber \\ && \times \Bigg(
    - 2 \sin\theta \, m_\Sigma \, (m_\Sigma+m_\Lambda) \, 
    \left( \frac{1}{t-m_\Sigma^2}+\frac{1}{u-m_\Sigma^2} \right) \nonumber \\ &&  \phantom{mm} {}
    - 2 \sin\theta 
  \Bigg)   \,.
  \label{eq:bornfeynnorm2}
\end{eqnarray}

Obviously we have pole terms and non-pole terms in (\ref{eq:bornfeynnorm1}) and (\ref{eq:bornfeynnorm2}). The 
respective p-wave projection is carried out by (\ref{eq:projE}), (\ref{eq:projM}). 
We denote the result for the pole terms by $K^{E/M}_{\rm Born}(s)$ and refrain from providing explicit expressions here. 
From (\ref{eq:bornfeynnorm1}), (\ref{eq:bornfeynnorm2}) one can immediately obtain 
\begin{eqnarray}
  \label{eq:detP0Born}
  P_{\rm Born}^M = P_{\rm Born}^E = -2 \, \frac{D \, F}{\sqrt{3} F_\pi^2} \,.
\end{eqnarray}

Note that the combination $\frac{1}{t-m_\Sigma^2}+\frac{1}{u-m_\Sigma^2}$ is a rational function in $s$, i.e.\ no square roots
show up. Thus there is no problem with the analytical continuation of the amplitude below its nominal threshold 
$s=(m_\Sigma+m_\Lambda)^2$. The same holds true for the combination 
$\frac{1}{p_{\rm c.m.} \, p_z} \left( \frac{1}{t-m_\Sigma^2}-\frac{1}{u-m_\Sigma^2} \right)$. Concerning the analytic structure of
the Born amplitudes in relation to electromagnetic form factors see also \cite{Meissner:1997ws}. 

From the NLO Lagrangian (\ref{eq:NLO1}) one obtains amplitudes $\sim \bar v_\Lambda u_\Sigma$ and 
$\sim \bar v_\Lambda  i \sigma_{\mu\nu}u_\Sigma p_+^\mu p_-^\nu$. The first amplitude 
does not contribute to $J=1$. The second one yields for the electric case, $\sigma=\lambda=+1/2$: 
\begin{eqnarray}
  && \bar v_\Lambda(-p_z,+1/2) i \sigma_{\mu\nu} u_\Sigma(p_z,+1/2) p_+^\mu p_-^\nu
  = \nonumber \\
  && \bar v_\Lambda(-p_z,+1/2) \, \gamma^3 \, u_\Sigma(p_z,+1/2) \, \frac{s \, p_{\rm c.m.}\cos\theta}{m_\Sigma+m_\Lambda}  \,,
  \label{eq:nlocontact1}
\end{eqnarray}
and for the magnetic case, $\sigma=-\lambda=+1/2$:
\begin{eqnarray}
  && \bar v_\Lambda(-p_z,-1/2) i \sigma_{\mu\nu} u_\Sigma(p_z,+1/2) p_+^\mu p_-^\nu
  = \nonumber \\
  && \bar v_\Lambda(-p_z,-1/2) \, \gamma^1 \, u_\Sigma(p_z,+1/2) \, p_{\rm c.m.}\sin\theta \, (m_\Sigma+m_\Lambda)  \,.
  \nonumber \\
  \label{eq:nlocontact2}
\end{eqnarray}
The overall coupling constant and flavor factor that multiplies these expressions to obtain the Feynman amplitude 
is $+4b_{10}/(\sqrt{3}F_\pi^2)$. 

As already spelled out, the electric part $\sim s$ is beyond our accuracy of NLO $\chi$PT. The magnetic part provides 
\begin{eqnarray}
  \label{eq:detNLOP0}
  P_{\rm NLO \, \chi PT}^M = \frac{4b_{10}}{\sqrt{3}F_\pi^2} \, (m_\Sigma+m_\Lambda)   \,.
\end{eqnarray}

Working with relativistic spin-3/2 Rarita-Schwinger fields is plagued by ambiguities how to deal with the spurious spin-1/2 
components. In the present context the interaction term $\sim h_A$ causes not only the proper exchange of spin-3/2 resonances, 
but induces an additional contact interaction. This unphysical contribution can be avoided by constructing interaction terms 
according to \cite{Pascalutsa:1999zz,Pascalutsa:2005nd} or \cite{Wies:2006rv}. 
The Pascalutsa prescription boils down to the replacement \cite{Pascalutsa:1999zz,Pascalutsa:2006up}
\begin{eqnarray}
  \label{eq:replace}
  T^\mu \to -\frac{1}{m_R} \, \epsilon^{\nu\mu\alpha\beta} \, \gamma_5 \, \gamma_\nu \, \partial_\alpha T_\beta
\end{eqnarray}
where $m_R$ denotes the resonance mass.
%\footnote{In \cite{Pascalutsa:2006up} also other prescriptions are provided. 
%The prescription used here can be found with different signs in the literature due to misprints and different conventions 
%how to define the Levi-Civita symbol. For the resonance-exchange diagrams this does not 
%matter, since the interaction term enters twice. We stick here to the conventions of \cite{pesschr}. 
%In this context, note, however, that in \cite{pesschr} the second equation in the text right after (3.68) has a wrong sign! 
%Note also that the spin-3/2 propagator given in (4.24) in 
%\cite{Pascalutsa:2006up} has a wrong sign in front of the $P_{21}$ term.} 
Strictly speaking this procedure induces an explicit flavor breaking, but such effects 
are anyway beyond leading order. In practice, we take the (average) mass of the $\Sigma^*$ resonance, 
$m_R:= m_{\Sigma^*}\approx 1.385\,$GeV. 

The spurious spin-1/2 components can only provide contact terms, i.e. polynomial terms, which do not have
a left-hand cut. Therefore they are completely irrelevant, if the polynomial part of the amplitude is determined 
anyway by matching the expression (\ref{eq:tmandel}) to $\chi$PT. This is the essence of (\ref{eq:fixpolyEM}). 
We will perform calculations with both interaction terms, the ``naive'' one given in (\ref{eq:baryonlagr}) and the 
``consistent'' interaction obtained by (\ref{eq:replace}). We will see that the pole terms remain unchanged.

Using the ``naive'' interaction term from (\ref{eq:baryonlagr}) the contributions from the exchange of $\Sigma^*$ 
decuplet baryon resonances are given by
\begin{eqnarray}
  && {\cal M}_{\rm res}^{\rm n} = \frac{h_A^2}{8 \sqrt{3} F_\pi^2} \, \nonumber \\ && \times \Bigg(
    \bar v_\Lambda u_\Sigma \, \frac{m_\Sigma+m_\Lambda}{12 m_{\Sigma^*}^2}  \, (t-u) \nonumber \\ && \phantom{m} {}
    - \bar v_\Lambda u_\Sigma \, F(s) \,
    \left( \frac{1}{t-m_{\Sigma^*}^2}-\frac{1}{u-m_{\Sigma^*}^2} \right) \nonumber \\ && \phantom{m} {}
    + \bar v_\Lambda \gamma^\mu k_\mu u_\Sigma \, \frac{1}{12 m_{\Sigma^*}^2}  \, \left( 
      - 2 m_{\Sigma^*}^2 - 2 m_{\Sigma^*}\, (m_\Sigma+m_\Lambda)      \right. \nonumber \\ && \hspace*{12.5em} \left. \phantom{)}
      + m_\Sigma^2 + m_\Lambda^2 +s - 6 m_\pi^2 \right)  \nonumber \\ && \phantom{m} {}
    + \bar v_\Lambda \gamma^\mu k_\mu u_\Sigma \, \frac12 \, G(s) \, 
    \left( \frac{1}{t-m_{\Sigma^*}^2}+\frac{1}{u-m_{\Sigma^*}^2} \right) \Bigg)
  \label{eq:resfeyn}
\end{eqnarray}
with
\begin{eqnarray}
  F(s) & := & \left( \frac{m_\Sigma+m_\Lambda}{2}+m_{\Sigma^*} \right) \, H_1(s) \nonumber \\
  && {} + \left( \frac{m_\Sigma+m_\Lambda}{2}-m_{\Sigma^*} \right) \, H_2  \,, \\
  G(s) & := & H_1(s)+H_2   \,, \\
  H_1(s) & := & \frac{m_\Sigma^2 + m_\Lambda^2-s}{2}    \nonumber \\
  && {} -\frac{(m_\Lambda^2+m_{\Sigma^*}^2-m_\pi^2)(m_\Sigma^2+m_{\Sigma^*}^2-m_\pi^2)}{4 m_{\Sigma^*}^2} \,,  \nonumber  \\
  H_2 & := & \frac13 \, \left( m_\Lambda + \frac{m_\Lambda^2+m_{\Sigma^*}^2-m_\pi^2}{2 m_{\Sigma^*}} \right)  \nonumber \\ 
  && \times 
  \left( m_\Sigma + \frac{m_\Sigma^2+m_{\Sigma^*}^2-m_\pi^2}{2 m_{\Sigma^*}} \right)  \,. \phantom{m}
  \label{eq:respoles-carlos2}
\end{eqnarray}

%Comparing the low-energy limit of our result (\ref{eq:resfeyn})
%to the results of \cite{Meissner:1997hn} provides a cross-check for our calculations. In fact it has been checked that our 
%result agrees with the one from equation (18) in \cite{Meissner:1997hn} concerning the contribution 
%$\sim 1/(m_{\rm decuplet}-m_{\rm octet})$. 
%We cannot reproduce the very last term in equation (18) in \cite{Meissner:1997hn} $\sim Z$. 
%
%We guess it is probably wrong. It should contain an extra factor
%$(1+2Z)/3$. Then, for our case $Z=-1/2$ all terms vanish except for the fist one $\sim 1/(m_{\rm decuplet}-m_{\rm octet})$.

As a a cross-check for our calculations we have calculated the nucleon- and Delta-exchange contributions to 
$p \bar n \to \pi^+ \pi^0$. 
They are related by large-$N_c$ relations; see appendix \ref{sec:Delta}. 

Only the contact terms change if one uses the ``consistent'' Pascalutsa interaction obtained by (\ref{eq:replace}):
\begin{eqnarray}
  &&  {\cal M}_{\rm res}^{\rm P} = \frac{h_A^2}{8 \sqrt{3} F_\pi^2} \, \nonumber \\ && \times \Bigg(
    \bar v_\Lambda u_\Sigma \, \frac{4 m_{\Sigma^*}+3 m_\Sigma+3 m_\Lambda}{12 m_{\Sigma^*}^2} \, (t-u) \nonumber \\ && \phantom{m} {}
    - \bar v_\Lambda u_\Sigma \, F(s) \,
    \left( \frac{1}{t-m_{\Sigma^*}^2}-\frac{1}{u-m_{\Sigma^*}^2} \right) \nonumber \\ && \phantom{m} {}
    + \bar v_\Lambda \gamma^\mu k_\mu u_\Sigma \, \frac{1}{12 m_{\Sigma^*}^2}  \, \nonumber \\ && \phantom{mm} \times 
    \left( -2 m_{\Sigma^*}^2 + 2 m_{\Sigma^*}\, (m_\Sigma+m_\Lambda) + 3m_\Sigma^2 + 3m_\Lambda^2 \right. 
    \nonumber \\ && \phantom{mmm} 
      \left. {} +4 m_\Sigma m_\Lambda -5s +2 m_\pi^2 \right)  
    \nonumber \\ && \phantom{m} {}
    + \bar v_\Lambda \gamma^\mu k_\mu u_\Sigma \, \frac12 \, G(s) \, 
    \left( \frac{1}{t-m_{\Sigma^*}^2}+\frac{1}{u-m_{\Sigma^*}^2} \right) \Bigg)  \,.  \phantom{mm}
  \label{eq:resfeynP}
\end{eqnarray}
We see that in (\ref{eq:resfeyn}) and (\ref{eq:resfeynP}) the pole terms agree for both prescriptions. 
We use these pole terms to obtain $K^{E/M}_{\rm res}(s)$. 
The polynomial contribution to the electric amplitude is given by 
\begin{eqnarray}
  P^E_{\rm res} & = & \frac{h_A^2}{24 \sqrt{3} F_\pi^2 \, m_{\Sigma^*}^2} \, \nonumber \\ && \phantom{m} \times 
  (m_{\Sigma^*}^2 + m_{\Sigma^*} \, (m_\Sigma+m_\Lambda) + m_\Sigma \, m_\Lambda) 
  \nonumber \\ && {} + {\cal O}(m_\pi^2, s)  \,.
  \label{eq:polyresel1}  
\end{eqnarray}
Only the subleading parts $\sim m_\pi^2, s$ are different for (\ref{eq:resfeyn}) and (\ref{eq:resfeynP}). We will 
neglect them in the following and just use 
\begin{eqnarray}
  P^E_{\rm res} & \approx & \frac{h_A^2}{24 \sqrt{3} F_\pi^2 \, m_{\Sigma^*}^2} \, \nonumber \\ && \times 
  (m_{\Sigma^*}^2 + m_{\Sigma^*} \, (m_\Sigma+m_\Lambda) + m_\Sigma \, m_\Lambda)   \,.
  \label{eq:polyresel2}  
\end{eqnarray}

Finally we need the low-energy limit of $K^{M}_{\rm res}(s)$ for the matching procedure spelled out 
in equation (\ref{eq:fixpolyEM}). Due to the left-hand cut structure of this amplitude there are some subtleties with this 
low-energy limit. This is discussed in detail in appendix \ref{sec:app-cutres}. The result of these considerations is
\begin{eqnarray}
  && K^M_{\rm res,low} = \frac{h_A^2}{24 \sqrt{3} F_\pi^2} \, \nonumber \\ && \times 
  \frac{(-m_{\Sigma^*}^2+4 m_{\Sigma^*} m_\Sigma - m_\Sigma^2) \, (m_{\Sigma^*} + m_\Sigma)}{m_{\Sigma^*}^2 \, (m_{\Sigma^*} - m_\Sigma)} \,.
  \label{eq:kmagreslowdefmain}
\end{eqnarray}

The formulae (\ref{eq:detP0Born}), (\ref{eq:detNLOP0}), (\ref{eq:polyresel2}), (\ref{eq:kmagreslowdefmain}) together with 
\begin{equation}
  K^{E/M}(s) = K^{E/M}_{\rm Born}(s) + K^{E/M}_{\rm res}(s)
  \label{eq:b+res}
\end{equation}
fully determine the input for (\ref{eq:tmandel}), (\ref{eq:fixpolyEM}).
Starting conceptually from octet $\chi$PT one can study the successive approximations of using 
a) ``Born'': only the Born terms, 
b) ``NLO'': Born plus NLO contact terms, and finally c) ``NLO+res'': the impact of including explicitly the decuplet exchange. 
Note that for the electric case there are no NLO corrections. Here we study ``Born'' and the addition of resonances.
To avoid a clutter of expressions we call the latter option also ``NLO+res''.

\section{Results}
\label{sec:res}

%As already spelled out we study the succession a) ``Born'', b) ``NLO'', 
%c) ``NLO+res''. Also note that there is no NLO part for the electric amplitudes. 

As a first step we fix the input parameters to the central values given in (\ref{eq:harange}) and (\ref{eq:b10range}), 
respectively. We address mainly two questions: How important is the exchange of decuplet resonances if their static part 
is already taken into account by NLO $\chi$PT? How strongly do the results depend on the cutoff $\Lambda$? 
We will show results for $\Lambda$ equal to 1 and 2 GeV, respectively. 

\begin{figure}[h!]
  \centering
  \begin{minipage}[c]{0.48\textwidth}
    \includegraphics[keepaspectratio,width=\textwidth]{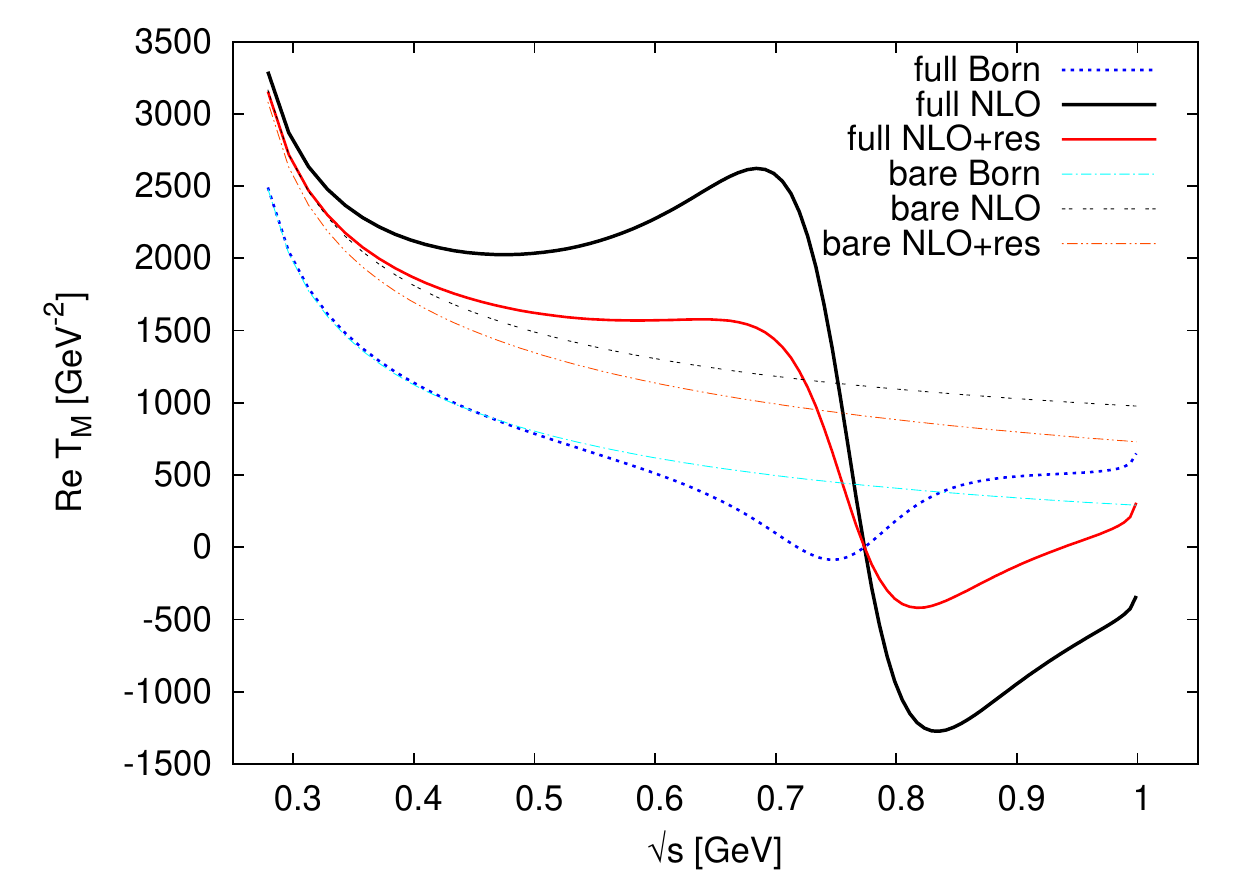}
  \end{minipage}  

  \begin{minipage}[c]{0.48\textwidth}  
    \includegraphics[keepaspectratio,width=\textwidth]{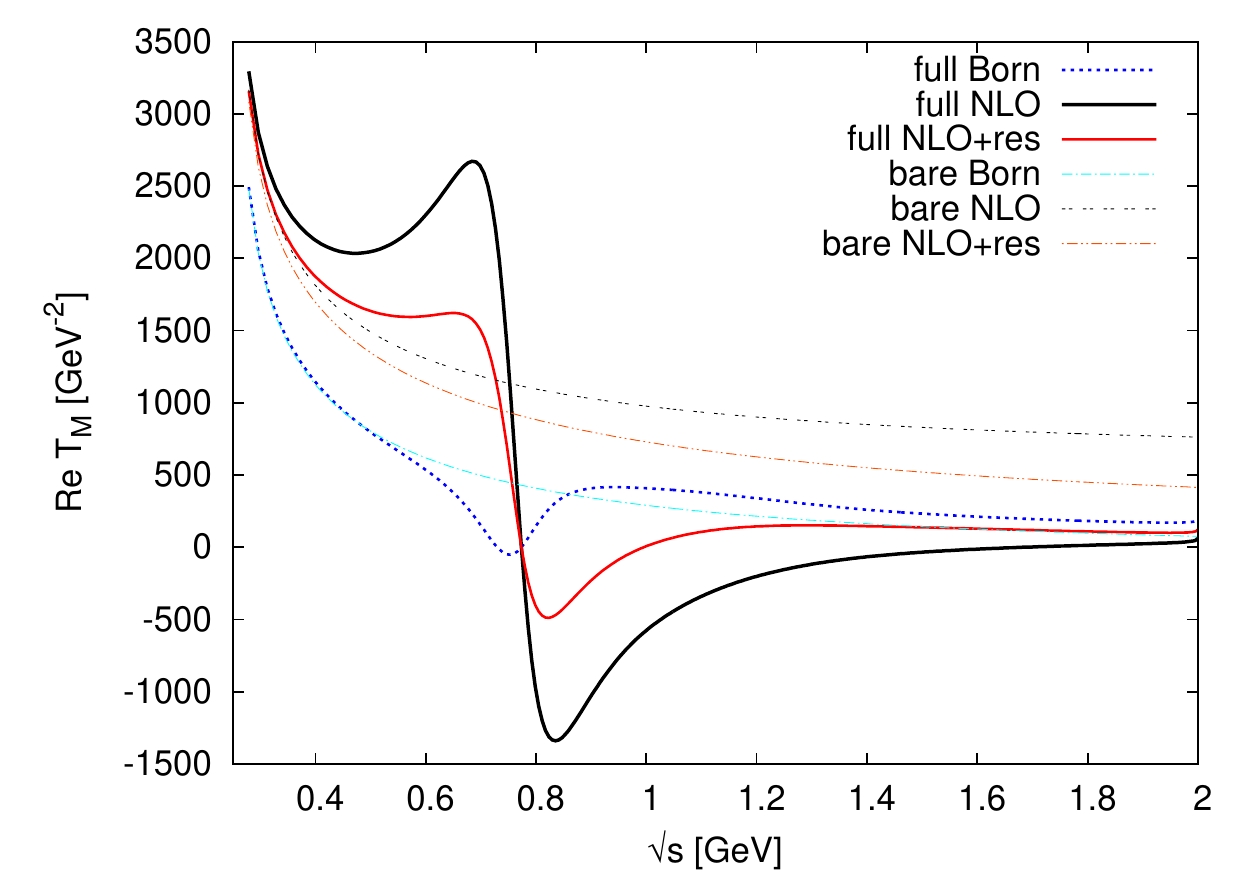}
  \end{minipage}   

  \begin{minipage}[c]{0.48\textwidth}  
    \includegraphics[keepaspectratio,width=\textwidth]{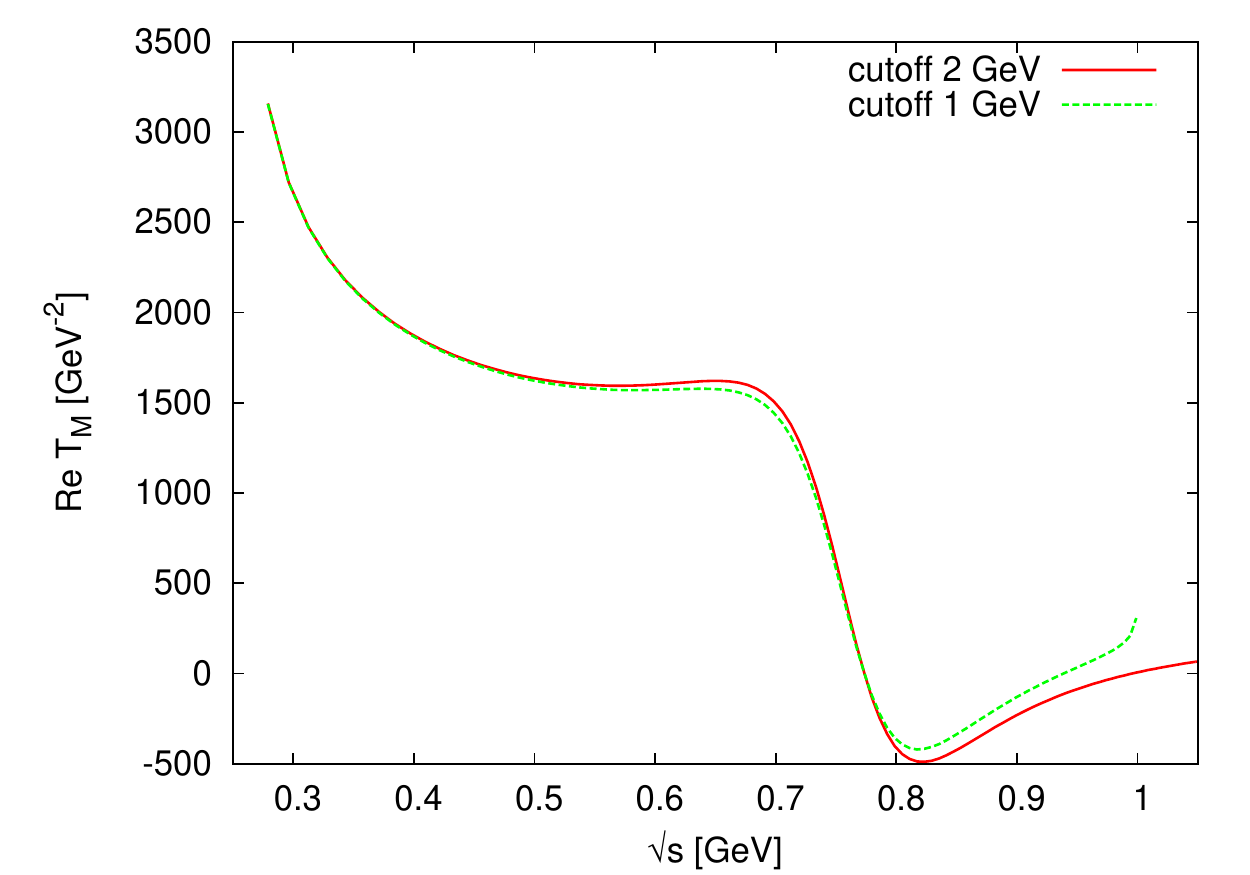}
  \end{minipage}   
  \caption{Real part of the magnetic scattering amplitude. 
    Top: $\Lambda = 1\,$GeV; middle: $\Lambda = 2\,$GeV; 
    bottom: Comparison of the complete result (``NLO+res'') for the two values of the cutoff $\Lambda$.
    The label ``bare'' denotes the input amplitude $K$, the label ``full'' the amplitude including pion rescattering. 
    For the non-color version we spell out the ordering of curves as they start out from the left. In the top and middle 
    panel from top to bottom one has 1.\ ``full NLO'', 2.\ ``full NLO+res'', 3.\ ``bare NLO'', 4.\ ``bare NLO+res'', 
    5.\ ``bare Born'', 6.\ ``full Born''. }
  \label{fig:resmagre}
\end{figure}

\begin{figure}[h!]
  \centering
  \begin{minipage}[c]{0.48\textwidth}
    \includegraphics[keepaspectratio,width=\textwidth]{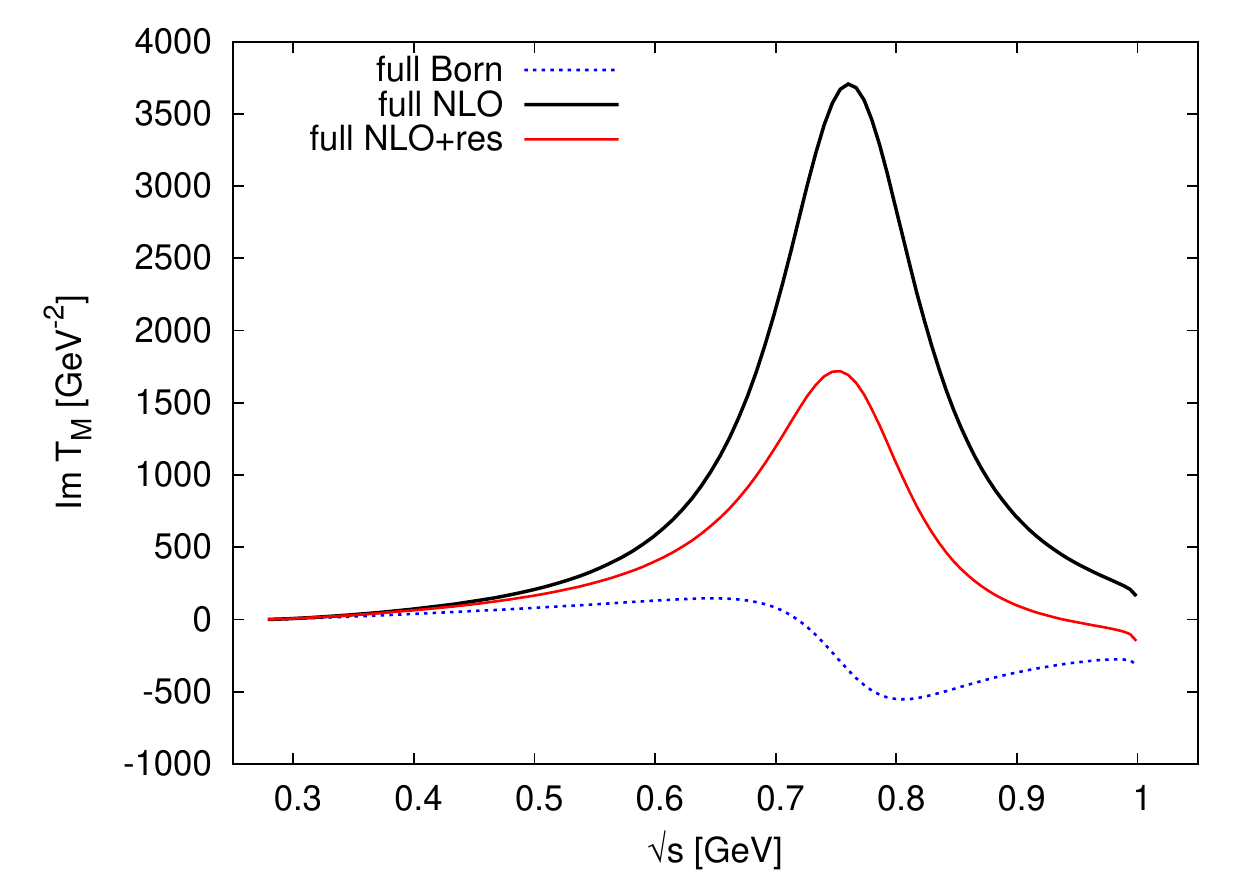}
  \end{minipage} 

  \begin{minipage}[c]{0.48\textwidth}  
    \includegraphics[keepaspectratio,width=\textwidth]{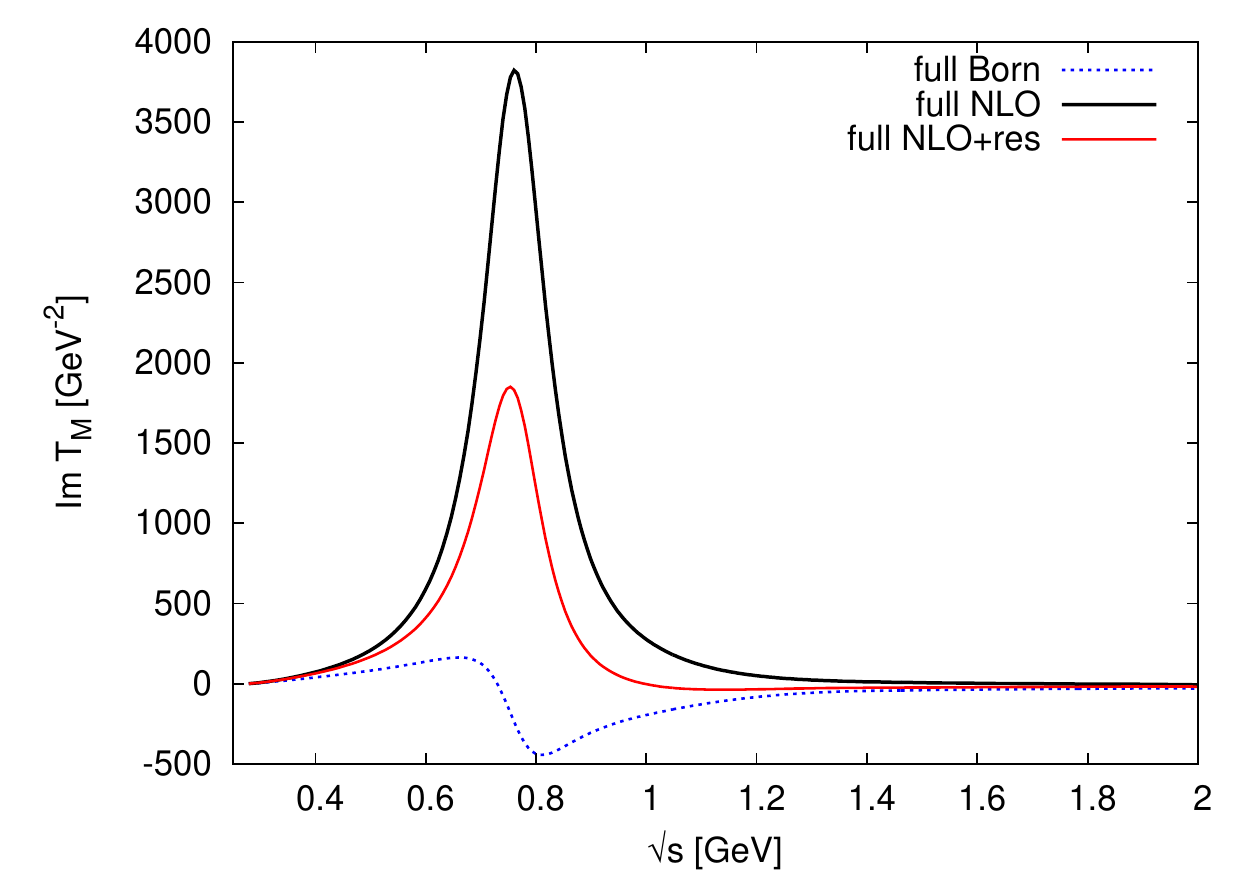}
  \end{minipage}   

  \begin{minipage}[c]{0.48\textwidth}  
    \includegraphics[keepaspectratio,width=\textwidth]{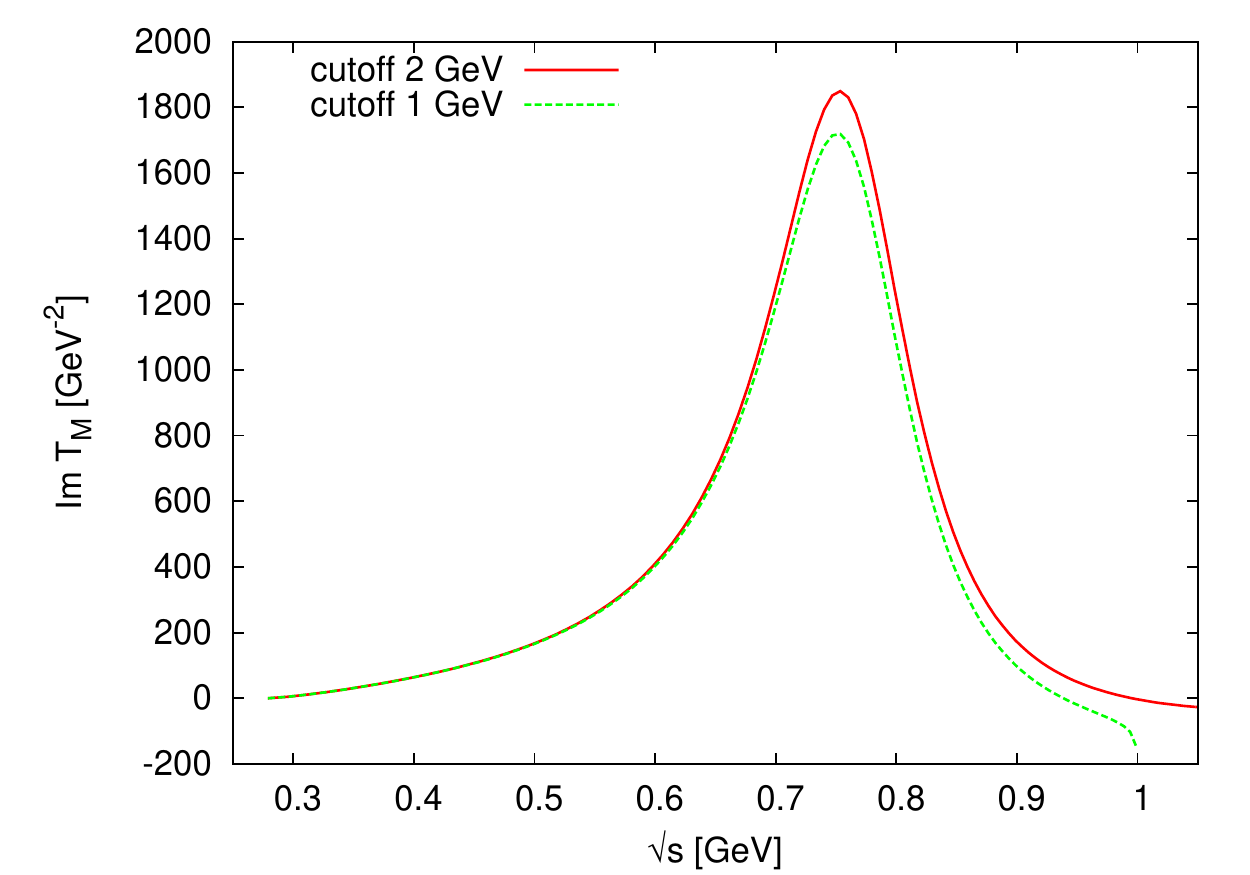}
  \end{minipage}   
  \caption{Imaginary part of the magnetic scattering amplitude. For the non-color version we spell out the ordering of curves.
    In the top and middle panel from top to bottom one has 1.\ ``full NLO'', 2.\ ``full NLO+res'', 3.\ ``full Born''. 
    Note that the bare amplitudes have no imaginary part in the displayed region. 
    See the caption of figure \ref{fig:resmagre} for more details.}
  \label{fig:resmagim}
\end{figure}

\begin{figure}[h!]
  \centering
  \begin{minipage}[c]{0.48\textwidth}
    \includegraphics[keepaspectratio,width=\textwidth]{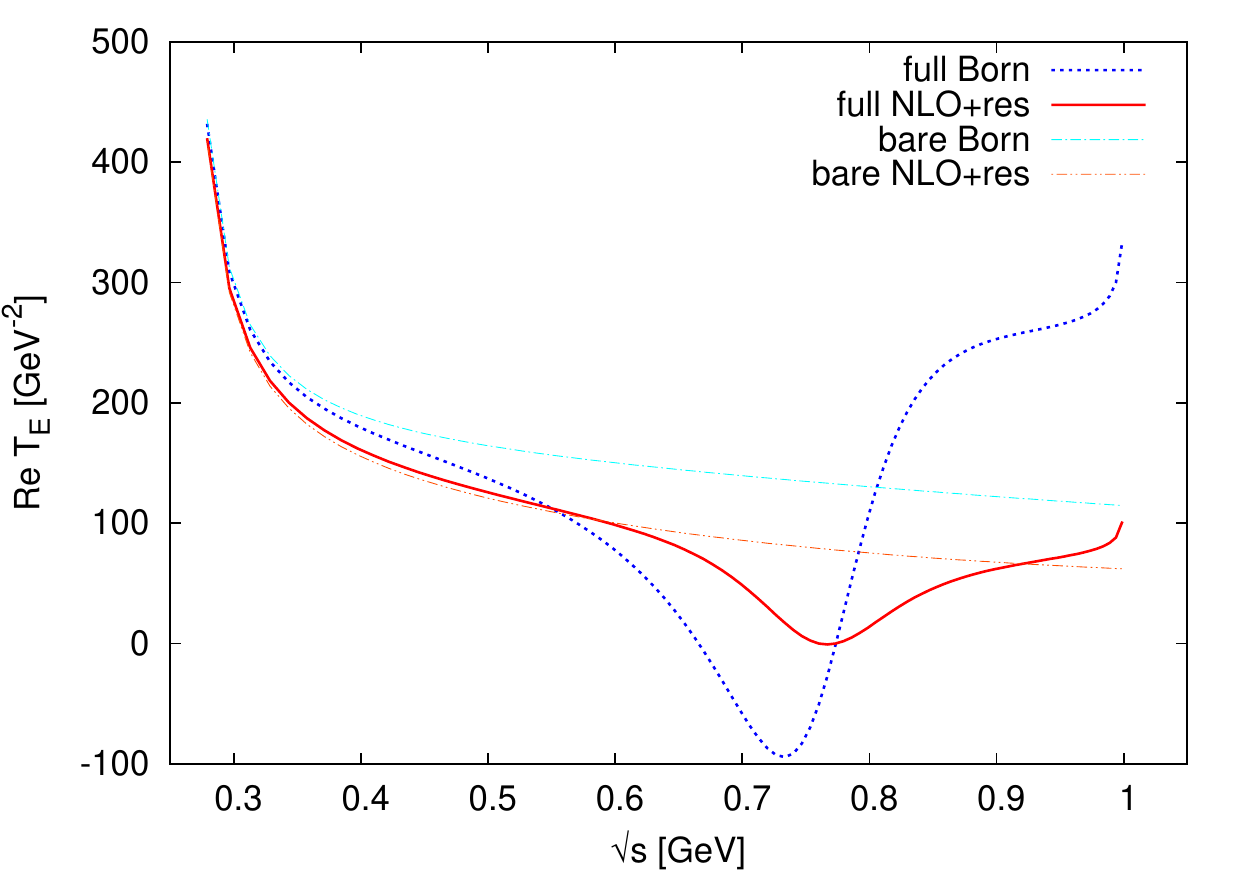}
  \end{minipage} 

  \begin{minipage}[c]{0.48\textwidth}  
    \includegraphics[keepaspectratio,width=\textwidth]{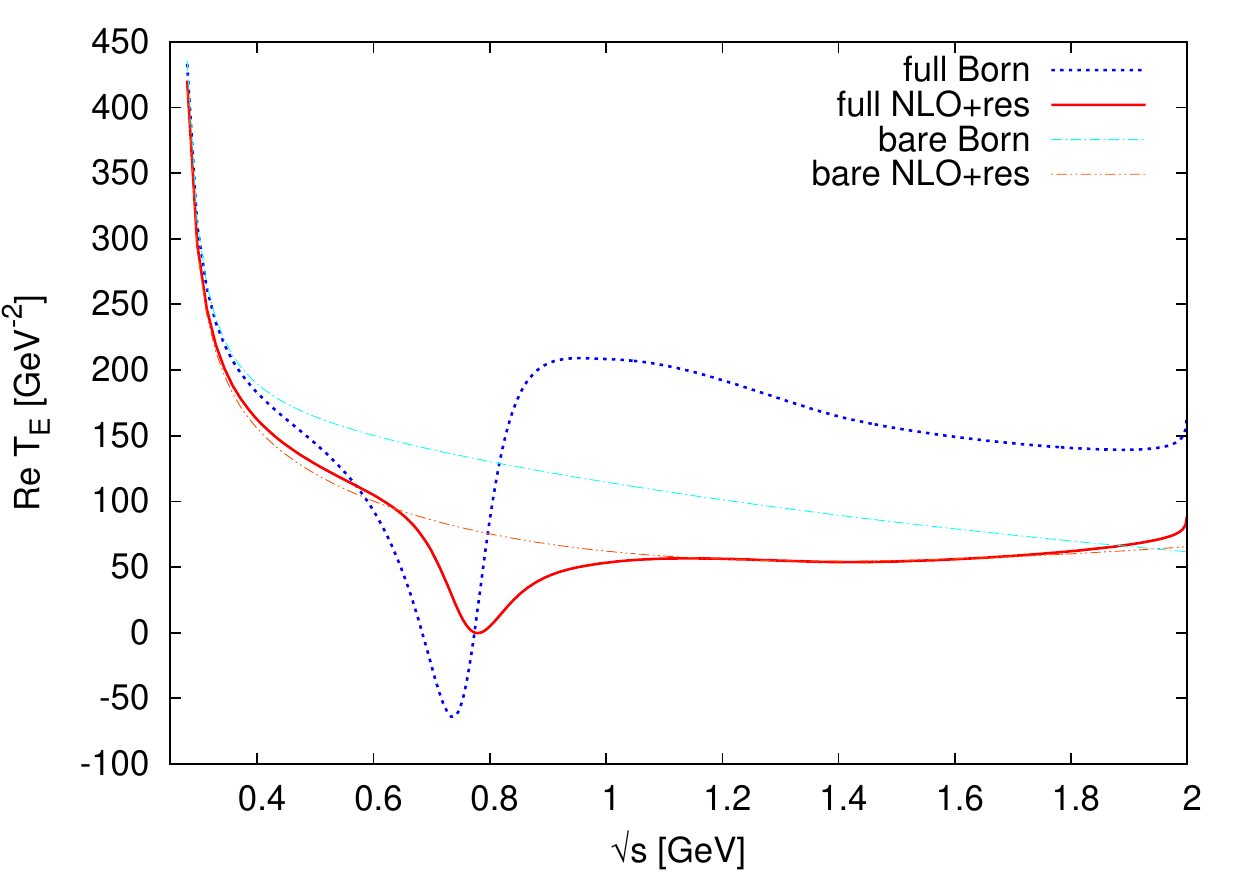}
  \end{minipage}   

  \begin{minipage}[c]{0.48\textwidth}  
    \includegraphics[keepaspectratio,width=\textwidth]{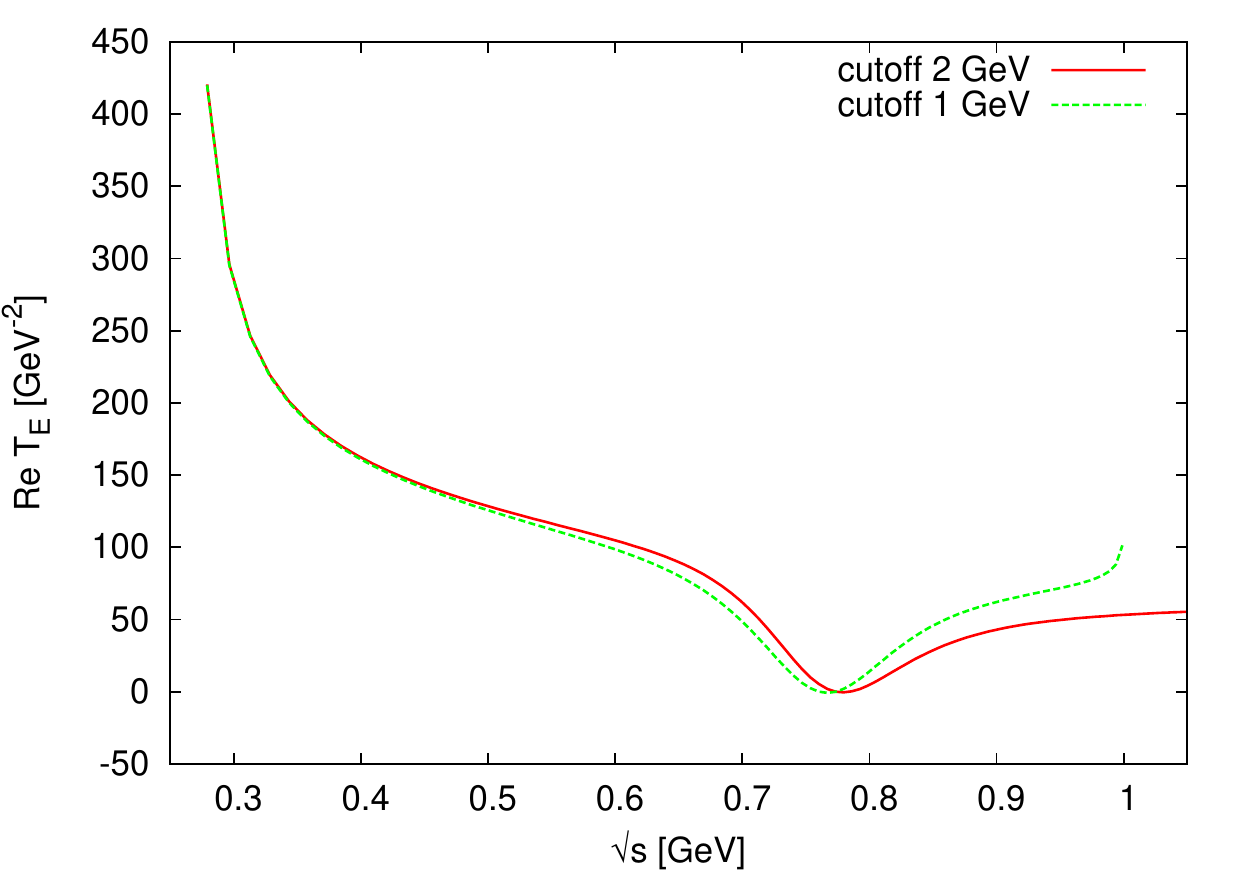}
  \end{minipage}   
  \caption{Real part of the electric scattering amplitude.
    For the non-color version we spell out the ordering of curves as they start out from the left. In the top and middle 
    panel from top to bottom one has 1.\ ``bare Born'', 2.\ ``full Born'', 3.\ ``full NLO+res'', 4.\ ``bare NLO+res''.
    See the caption of figure \ref{fig:resmagre} for more details.}
  \label{fig:reselre}
\end{figure}

\begin{figure}[h!]
  \centering
  \begin{minipage}[c]{0.48\textwidth}
    \includegraphics[keepaspectratio,width=\textwidth]{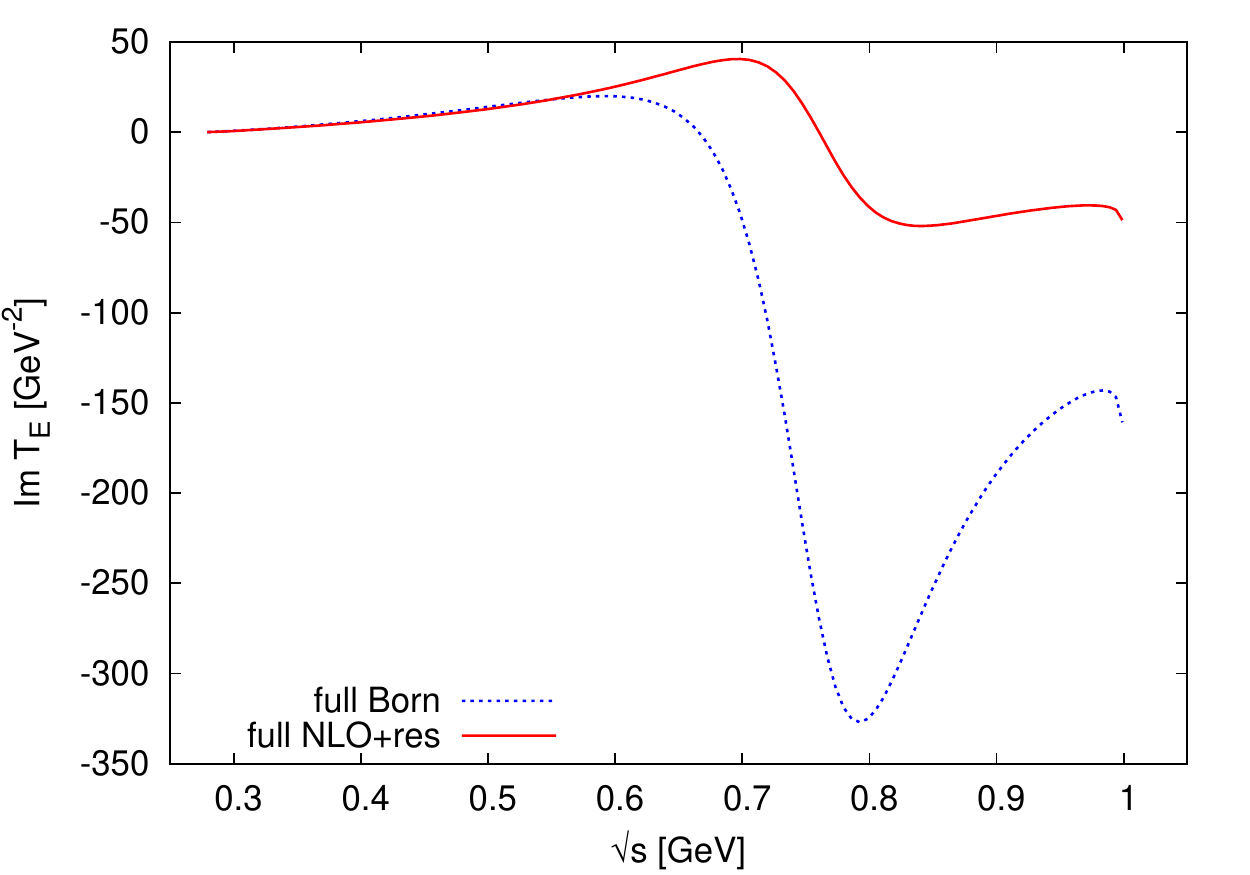}
  \end{minipage} 

  \begin{minipage}[c]{0.48\textwidth}  
    \includegraphics[keepaspectratio,width=\textwidth]{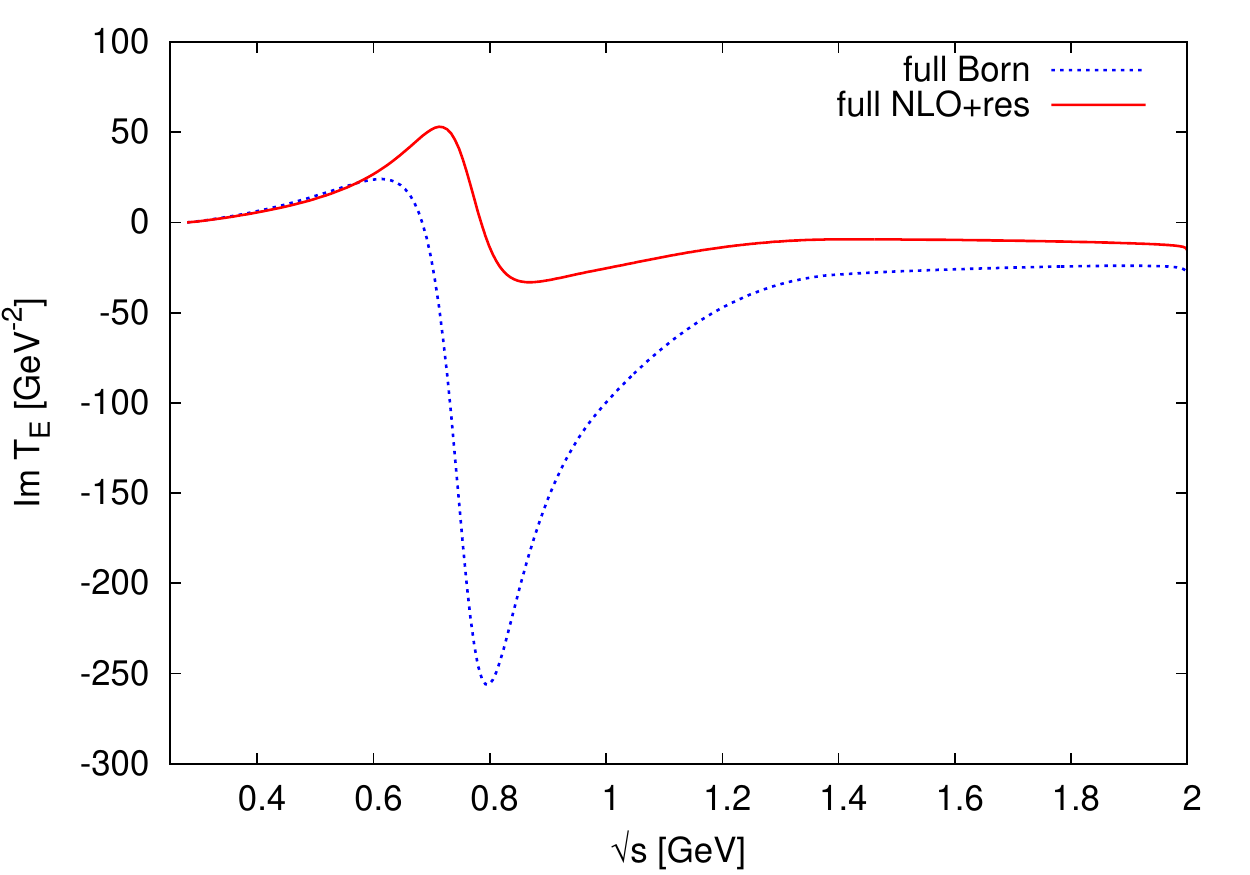}
  \end{minipage}   

  \begin{minipage}[c]{0.48\textwidth}  
    \includegraphics[keepaspectratio,width=\textwidth]{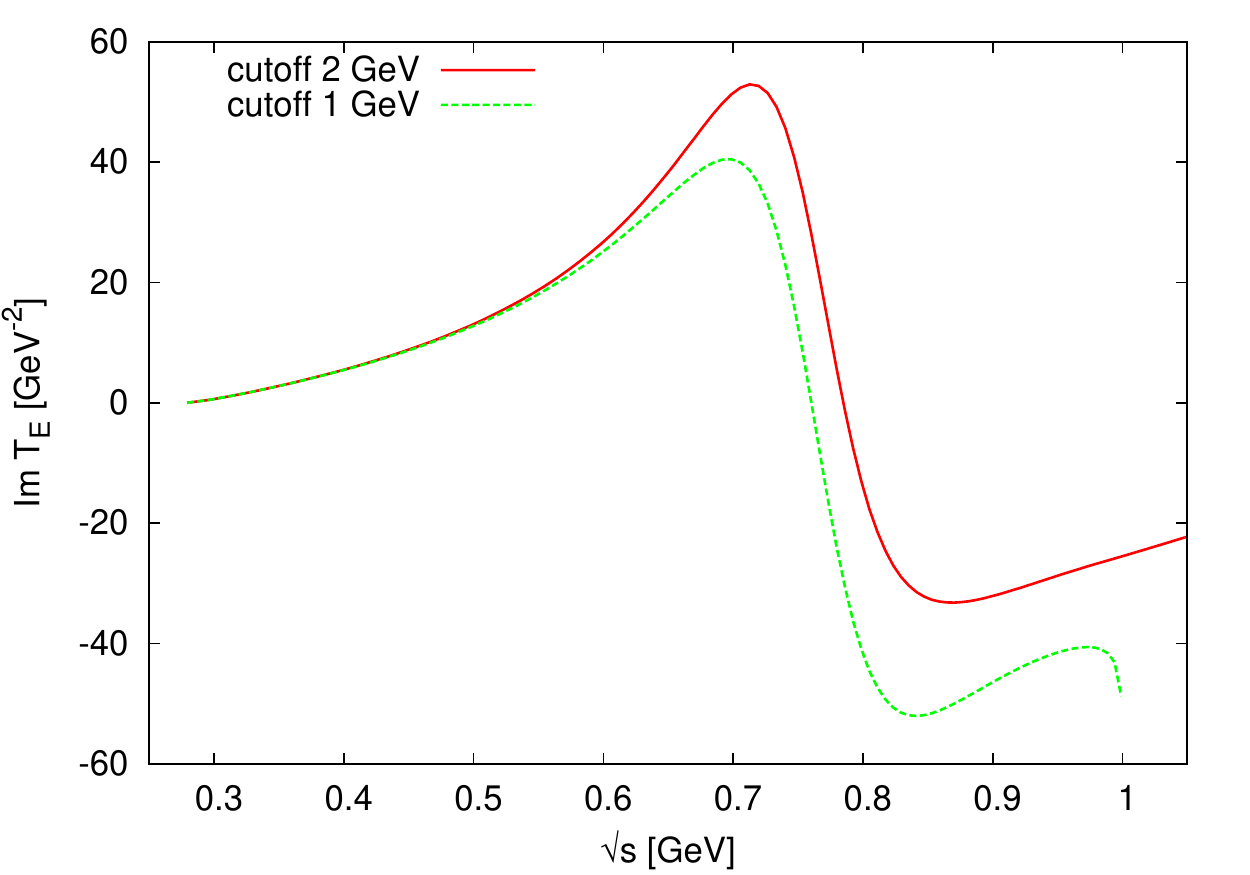}
  \end{minipage}   
  \caption{Imaginary part of the electric scattering amplitude. 
    See the caption of figure \ref{fig:resmagre} for more details.}
  \label{fig:reselim}
\end{figure}

Figures \ref{fig:resmagre}-\ref{fig:reselim} show real and imaginary part of 
the formal (sub-threshold) electric and magnetic scattering amplitudes. 
Obviously the pion rescattering drastically reshapes 
the amplitudes providing the expected structure around the rho meson. Only at very low energies the impact of pion rescattering 
is negligible. From the magnetic sector we observe that the 
NLO term qualitatively changes all the results as compared to the pure Born terms. The explicit resonance terms do not change 
the qualitative picture any more, but ``damp'' to some extent the structures emerging at NLO. In the electric 
sector we observe also that the resonances matter. The impact of the variation in the cutoff, however, is in general 
rather small. This is very satisfying given that a $\chi$PT input loses its validity at higher energies. 

From studying the scattering amplitudes alone one cannot tell easily how strongly the form factors are influenced. 
Table \ref{tab:kaprad} documents the results from the unsubtracted dispersion relations (\ref{eq:dispbasicunsubtrkappa}) 
and (\ref{eq:dispbasicunsubtrnull}) for the multipole moments and from the subtracted dispersion relations (\ref{eq:dispbasic}) 
for the radii (\ref{eq:defradiusEME}), (\ref{eq:defradiusM}) using the central values of $h_A$ and $b_{10}$.
\begin{table*}[h!]
  \centering
  \begin{tabular}{|c|c|c|c|c|c|}
    \hline
    $\Lambda$ [GeV] & quantity & Born & NLO & NLO+res & $\chi$PT \\  \hline \hline
    $1$ & $G_M(0)$ & $-0.438$ & $5.55$ & $2.58$ & $1.98$ (exp.) \\  \hline 
    $2$ &          & $-0.65$ &  $5.98$ & $2.66$ & \\  \hline \hline
    $1$ & $\langle r_M^2 \rangle$ [GeV$^{-2}$] & $0.453$ & $33.7$ & $17.9$ & $18.6$ \\  \hline 
    $2$ &                                    & $0.613$  & $35.2$ & $18.8$ & \\  \hline \hline
    $1$ & $G_E(0)$ & $-0.432$ & - & $0.0026$ & $0$  \\  \hline 
    $2$ &          & $-0.562$ & - & $-0.031$ & \\  \hline \hline
    $1$ & $\langle r_E^2 \rangle$ [GeV$^{-2}$] & $-3.13$ & - & $0.866$ & $0.773$ \\  \hline 
    $2$ &                                    & $-2.91$ & - & $1.044$ & \\  \hline 
  \end{tabular}
  \caption{Comparison to $\chi$PT \cite{Kubis:2000aa} using $h_A = 2.3$, $b_{10}=1.1\,$GeV$^{-1}$.}
  \label{tab:kaprad}
\end{table*}
In general one observes that the Born terms alone are insufficient to produce reasonable results. The inclusion of the NLO 
term and/or the decuplet-resonance exchange improves the picture significantly --- signs and orders of magnitude come out 
correctly. 
Interestingly even the unsubtracted dispersion relations produce quite reasonable results. In particular in the 
electric sector the resonance exchange has the potential to cancel the Born contribution such that essentially 
the vanishing of the electric charge is achieved. 
In most of the cases varying the cutoff provides changes on the level of 10\% at most. 
Thus the dispersive representation is most sensitive to the low-energy regime. This is an encouraging result given the 
$\chi$PT input and the fact that not considered inelasticities like kaon-antikaon come into play at around 1 GeV. 

\begin{table}[h!]
  \centering
  \begin{tabular}{|c|c|c|c|c|}
    \hline
    $b_{10}$ & quantity & NLO & NLO+res & $\chi$PT \\  \hline \hline
    $0.85$ & $G_M(0)$ & $4.47$ & $1.15$ & $1.98$ (exp.) \\  \hline 
    $1.35$ &          & $7.49$ & $4.17$ & \\  \hline \hline
    $0.85$ & $\langle r_M^2 \rangle$ [GeV$^{-2}$] & $27.4$ & $10.9$ & $18.6$ \\  \hline 
    $1.35$ &                                     & $43.1$ & $26.7$ & \\  \hline 
  \end{tabular}
  \caption{Comparison to $\chi$PT \cite{Kubis:2000aa} using $\Lambda = 2 \,$GeV, $h_A = 2.3$ and varying the 
    value for $b_{10}$ (in units of GeV$^{-1}$).}
  \label{tab:kapradb10}
\end{table}
After having convinced ourselves that the variation of the cutoff produces only moderate changes we keep $\Lambda = 2\,$GeV 
fixed and explore the impact of variations of the other two input parameters. 
In table \ref{tab:kapradb10} we explore the changes of the low-energy quantities if the value of $b_{10}$ is varied 
according to (\ref{eq:b10range}). Since the electric sector is independent of $b_{10}$ we restrict ourselves 
to the magnetic quantities. 
The conclusions to be drawn from inspecting table \ref{tab:kapradb10} are: 
One needs the decuplet, only then one obtains reasonable values for the magnetic radius. 
Interestingly even the unsubtracted dispersion relation works not too badly. However, the uncertainty related to $b_{10}$ is 
sizable. Results change by a factor of 2 or more. 
Clearly a much better knowledge of $b_{10}$ is mandatory to improve on the predictions in the magnetic 
sector. 

Table \ref{tab:kapradha} displays the consequences of the variation in $h_A$ according to (\ref{eq:harange}).
\begin{table}[h!]
  \centering
  \begin{tabular}{|c|c|c|c|}
    \hline
    quantity & $h_A = 2.2$ & $h_A = 2.4$ & $\chi$PT \\  \hline \hline
    $G_M(0)$ & $2.94$ & $2.36$ & $1.98$ (exp.) \\  \hline 
    $\langle r_M^2 \rangle$ [GeV$^{-2}$] & $20.2$ & $17.3$ & $18.6$ \\  \hline 
    $G_E(0)$ & $-0.076$ & $0.016$ & $0$  \\  \hline 
    $\langle r_E^2 \rangle$ [GeV$^{-2}$] & $0.708$ & $1.40$ & $0.773$ \\  \hline 
  \end{tabular}
  \caption{Comparison of the full calculation ``NLO+res'' to $\chi$PT \cite{Kubis:2000aa} using $\Lambda = 2 \,$GeV, 
    $b_{10}=1.1\,$GeV$^{-1}$ and varying the value for $h_A$.}
  \label{tab:kapradha}
\end{table}
In the magnetic sector the changes caused by variations in $h_A$ are moderate. Thus once one has achieved a better handle
on $b_{10}$, then satisfying predictive power for the magnetic sector can be achieved. 
In other words, a measurement of the magnetic transition radius, e.g.\ at FAIR, would pin down $b_{10}$ and drastically decrease
the uncertainties of the low-energy magnetic transition form factor.  

The electric sector is independent 
of $b_{10}$. Table \ref{tab:kapradha} shows that for reasonable values of $h_A$ the value of $G_E(0)$ can even be 
fine-tuned to zero. The smallness of the electric radius as predicted in \cite{Kubis:2000aa} is qualitatively reproduced. 

\begin{table}[h!]
  \centering
  \begin{tabular}{|c|c|c|c|c|}
    \hline
    quantity & Born & NLO & NLO+res & $\chi$PT \\  \hline \hline
    $G_M(0)$ & $-0.648$ & $5.74$ & $2.65$ & $1.98$ (exp.) \\  \hline 
    $\langle r_M^2 \rangle$ [GeV$^{-2}$] & $0.613$ & $34.0$ & $18.7$ & $18.6$ \\  \hline 
    $G_E(0)$ & $-0.562$ & - & $-0.068$ & $0$  \\  \hline 
    $\langle r_E^2 \rangle$ [GeV$^{-2}$] & $-2.907$ & - & $0.774$ & $0.773$ \\  \hline 
  \end{tabular}
  \caption{Comparison to $\chi$PT \cite{Kubis:2000aa} using $\Lambda = 2 \,$GeV, $h_A = 2.22$, $b_{10}=1.06 \,$GeV$^{-1}$.}
  \label{tab:kapradft}
\end{table}
Obviously by just tuning the parameters in reasonable ranges 
all electric and magnetic low-energy quantities can be reproduced  ---
not all at the same time, but one would not expect the unsubtracted dispersion relations to hold exactly. 
To illustrate this further we tune $h_A$ and $b_{10}$ such that the electric and magnetic radii are essentially reproduced. 
The results are shown in table \ref{tab:kapradft}.
In particular we needed only a little change in $h_A$ to fine-tune the comparatively rather small electric radius. 
The point here is that the Born and resonance exchange contributions nearly cancel each other as can be seen from the 
comparison of the pure Born and the complete result for the electric radius in both 
tables \ref{tab:kaprad} and \ref{tab:kapradft}. We note in passing that this is qualitatively in line with the 
large-$N_c$ considerations discussed in appendix \ref{sec:Delta}. 

We will not explore at all the impact 
of a variation of the other input on our calculations. The parameters $F$ and $D$ are better constrained than $h_A$. 
Given our quite sizable uncertainties there is no point in exploring in this first paper the consequences from the 
differences in the 
pion phase shift as provided in \cite{GarciaMartin:2011cn} or \cite{Colangelo:2001df}, respectively. For the results we have 
utilized the phase shift from \cite{GarciaMartin:2011cn}. The same remark applies to the 
differences between the pion form factor and the Omn\`es function at larger energies \cite{Hanhart:2012wi}. All these 
uncertainties can be explored if the parameters $b_{10}$ and $h_A$ are better under control and/or if a $\chi$PT calculation 
for the hyperon-pion amplitudes beyond NLO is used.

\begin{figure}[h!]
  \centering
  \begin{minipage}[c]{0.48\textwidth}  
    \includegraphics[keepaspectratio,width=\textwidth]{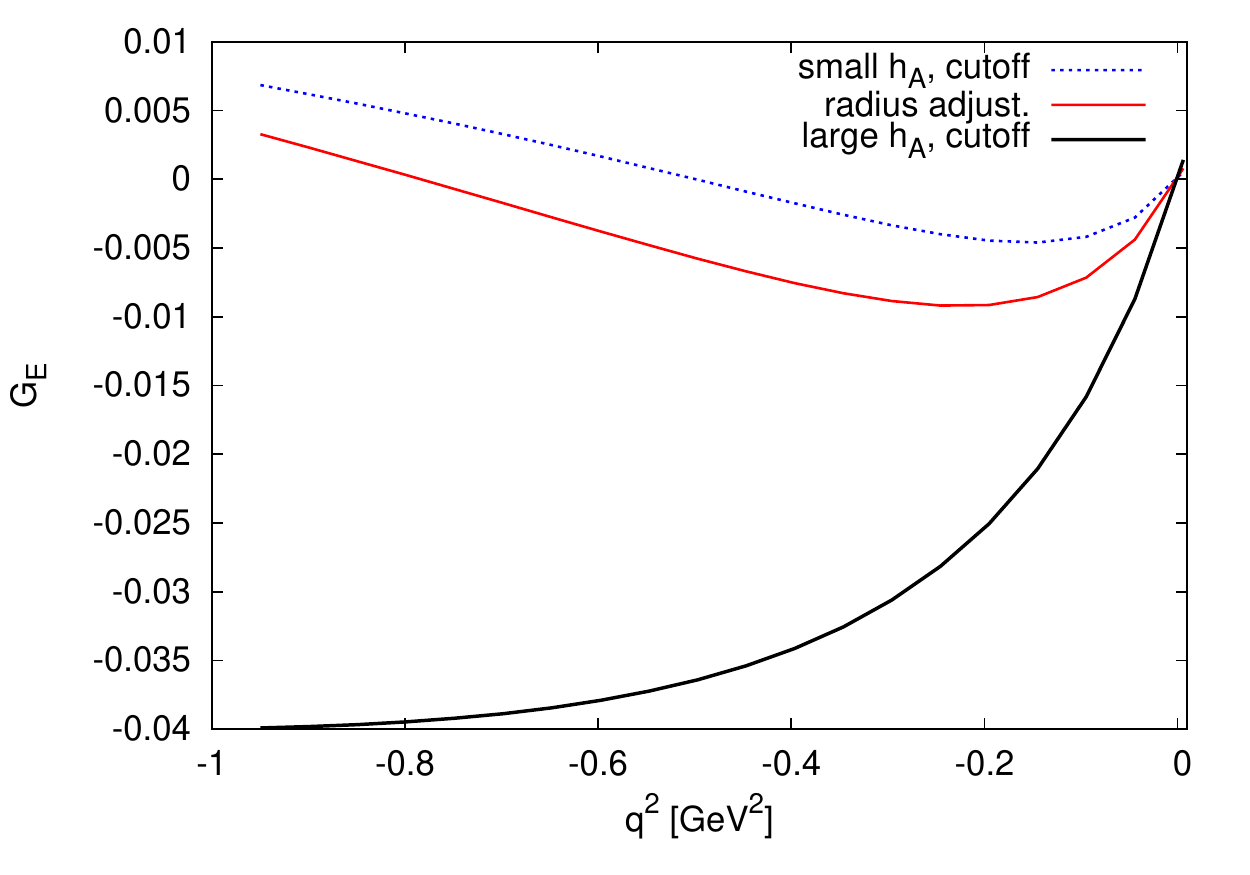}
  \end{minipage}   
  \caption{The electric transition form factor obtained from the subtracted 
    dispersion relation (\ref{eq:dispbasic}). The curve labeled by ``radius adjust.'' is obtained with the parameter
    values of table \ref{tab:kapradft}. The label ``small/large $h_A$, cutoff'' refers to the lower/upper value of the 
    range (\ref{eq:harange}) and to $\Lambda = 1/2\,$GeV. See main text for details.
    For the non-color version we spell out the ordering of curves. In the space-like region one has from top to bottom 
    the same ordering as in the caption in the figure. }
  \label{fig:resGE1}
\end{figure}
We use the subtracted dispersion relations (\ref{eq:dispbasic}) to determine the form factors in the 
region $q^2 < (m_\Sigma-m_\Lambda)^2$. The electric transition form factor is shown in figure \ref{fig:resGE1}. 
By varying $h_A$ in the range (\ref{eq:harange}) and the cutoff $\Lambda$ between 1 and 2 GeV we created a family of curves. 
In figure \ref{fig:resGE1} we show the respective highest and lowest curve. In addition we show the curve determined with the 
parameter values of table \ref{tab:kapradft}, which have been adjusted to the central value of the electric radius as obtained 
in \cite{Kubis:2000aa}. The main conclusion from figure \ref{fig:resGE1} is that the electric transition form factor remains 
quite small over a large range of $q^2$. This is somewhat different from the result of \cite{Kubis:2000aa} where a larger 
curvature and therefore a larger variation with $q^2$ has been found. 
Note, however, that this curvature is not obtained from pure $\chi$PT but from introducing an additional lagrangian for 
vector mesons into the framework. Naturally this is associated with some model uncertainties that are hard to quantify. 
In our approach the $\rho$-meson --- the only relevant vector meson in the isovector channel --- is included 
by dispersion theory. On the other hand, given the restriction of our $\chi$PT input to NLO, we do not want to claim that 
we have our uncertainties fully under control. But the results from tables \ref{tab:kaprad}-\ref{tab:kapradft} are encouraging.

\begin{figure}[h!]
  \centering
  \begin{minipage}[c]{0.48\textwidth}  
    \includegraphics[keepaspectratio,width=\textwidth]{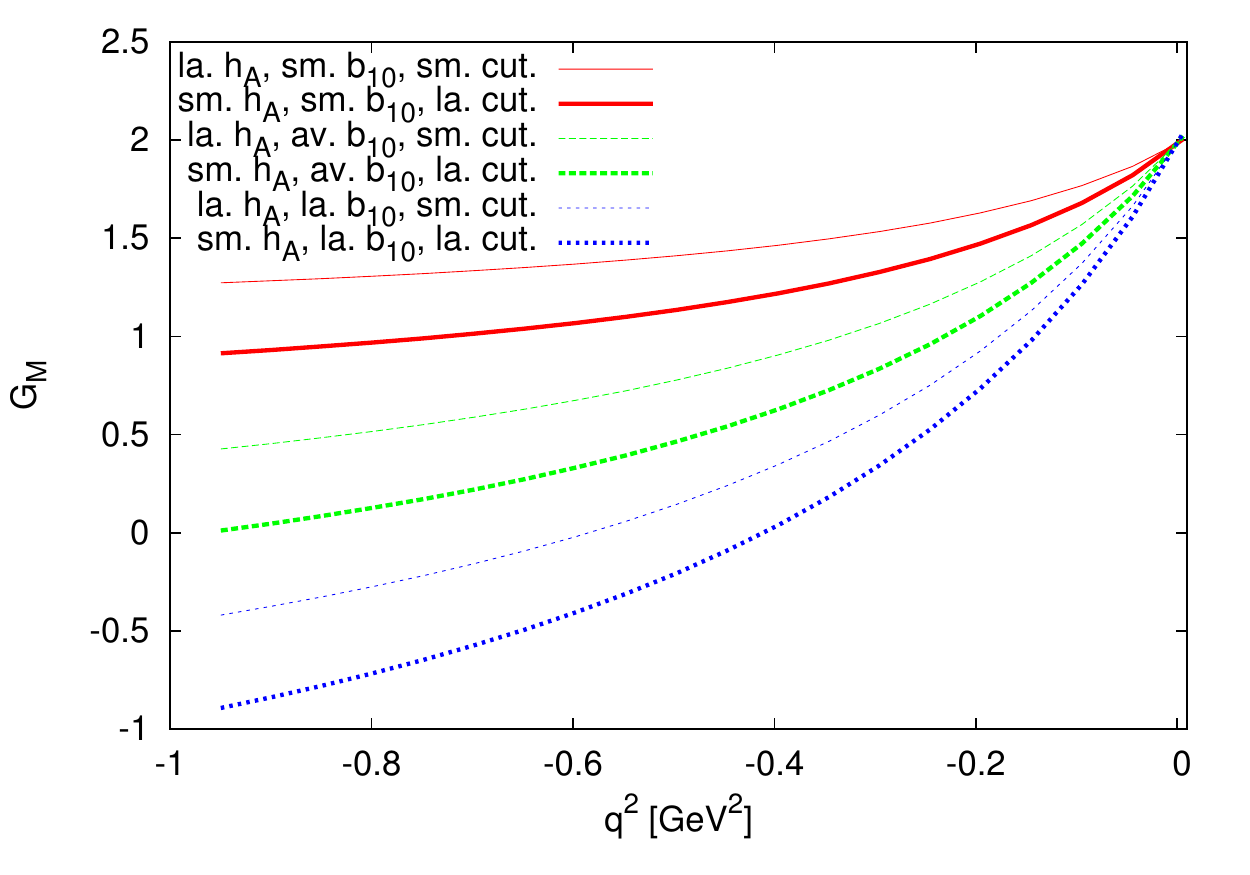}
  \end{minipage}   
  \caption{The magnetic transition form factor obtained from the subtracted 
    dispersion relations (\ref{eq:dispbasic}). The label ``sm./la.\ $h_A$'' refers to the lower/upper value of the 
    range (\ref{eq:harange}). The label ``sm./av./la.\ $b_{10}$'' refers to the lower/medium/upper value of the 
    range (\ref{eq:b10range}). The label ``sm./la.\ cut.'' refers to $\Lambda = 1/2\,$GeV.
    For the non-color version we spell out the ordering of curves. In the space-like region one has from top to bottom 
    the same ordering as in the caption in the figure. }
  \label{fig:resGM2}
\end{figure}
The magnetic transition form factor is presented in figure \ref{fig:resGM2}. Concerning the variation in $h_A$ and 
$\Lambda$ we show again only the curves that encompass the whole respective family of curves. 
Obviously the value for $b_{10}$ has the largest impact on the magnetic transition form factor. 
A better knowledge of $b_{10}$ would significantly decrease the uncertainty.

\section{Further discussion, summary and outlook}

It is worth to compare a direct $\chi$PT calculation to our framework, which combines dispersion theory with purely 
hadronic $\chi$PT. A relativistic $\chi$PT calculation up to (including) order $Q^4$, i.e.\ a full one-loop calculation, 
has been presented in \cite{Kubis:2000aa,KubisPhD}. Our dispersive approach includes automatically all the 
$\chi$PT one-loop diagrams where the virtual photon couples to the pion. However, it does not include the loop diagrams 
where the photon 
couples to kaons or baryons. On the other hand, from the point of view of a dispersive representation the contribution of the 
diagrams with the coupling of the photon to kaons or baryons is suppressed at low energies, in particular for a subtracted 
dispersion relation. Thus, such diagrams might contribute to the absolute size of a form factor, but not much to the energy 
variation of the form factor. Therefore we expect that using the experimental values for the form factors at the photon point 
but extracting radii and general shape from the (subtracted) dispersive representation should effectively include the 
dominant physics contained in a one-loop $\chi$PT calculation. Dispersion theory goes even beyond one-loop by including 
pion rescattering to all orders. Thus the dynamical effects of the $\rho$-meson are automatically included, not only in a static
approximation via low-energy constants or by adding a vector-meson lagrangian to $\chi$PT, which induced to some extent a 
model dependence. Of course, one future improvement of the present formalism could be to include the kaon-antikaon inelasticity 
and explore its impact on the shape of the form factors. Strictly speaking one should also include on the same footing the 
four-pion inelasticity, but maybe the corresponding three-loop diagrams are less important. 

There is a second aspect where our combined framework might be superior to a pure $\chi$PT calculation. 
There the decuplet states are not included as active degrees of freedom since they are not degenerate with the octet in the 
chiral limit. The decuplet appears indirectly in a static version by influencing the low-energy constants. 
Our results suggest that the explicit inclusion of dynamical decuplet states might be important, 
qualitatively in line with, e.g., \cite{Lutz:2001yb,Pascalutsa:2005nd,Pascalutsa:2006up,Ledwig:2014rfa}. 
Of course, the inclusion of a dynamical decuplet can also be performed in a $\chi$PT framework, but the whole development is in 
an infant stage. Not even the full NLO 
lagrangian for relativistic three-flavor octet+decuplet baryon $\chi$PT has been formulated up to now. 
This deficiency will also concern our framework when we will apply it to the decuplet-to-octet transition form factors. 
For the present work, where the external states are octet hyperons, the inclusion of the decuplet intermediate states was 
very straightforward. Since the pionic loops are generated by the dispersive representation and not by an explicit loop 
calculation, there is no ambiguity related to the renormalization of the 
loop diagrams \cite{Kubis:2000aa,Lutz:2001yb,Pascalutsa:2005nd,Pascalutsa:2006up,Scherer:2012xha,Ledwig:2014rfa}. 

For our hadronic input for the pion-hyperon scattering amplitudes we have used the language of three-flavor $\chi$PT. 
Conceptually this seems to be at odds with our dispersive setup where we include only the pions as intermediate states 
and not also the kaons. As already pointed out the influence of the kaon inelasticity starts at rather high energies, 
$(2 m_K)^2 \approx 1\,$GeV$^2$. In principle, we could have formulated our whole framework in a language with two light flavors.
There, pions would couple to the isosinglet $\Lambda$ and the isotriplet $\Sigma$ and $\Sigma^*$ states. 
Without the notion of a 
three-flavor chiral limit the mass difference between $\Lambda$ and $\Sigma$ would not vanish, but one can assume nonetheless 
that the mass difference is a small quantity --- similar in spirit to the assumption that the mass difference between 
$\Sigma$ and $\Sigma^*$ constitutes another small quantity, which does not vanish in the chiral limit. 
In such a framework the coupling constants for $\Sigma^*$-$\Lambda$-$\pi$
and $\Sigma^*$-$\Sigma$-$\pi$ would be independent from each other. But both can be determined from the $\Sigma^*$ partial decay
widths. This is what we essentially did anyway. In this framework with two light flavors those coupling constants would be 
independent of the $\Delta$-$N$-$\pi$ coupling constant. For our numerics we did not use such a three-flavor relation.
The only input that we used from a three-flavor analysis are the values for $D$ and $F$, i.e.\ the coupling constants for 
$\Sigma$-$\Lambda$-$\pi$ and $\Sigma$-$\Sigma$-$\pi$, respectively. In a two-flavor framework these constants would not be 
related to $g_A$, i.e.\ to the $N$-$N$-$\pi$ coupling constant. In our present analysis we use $D$ and $F$ just as input. 
Of course, their sizes are determined from a three-flavor formalism, but the flavor breaking is relatively well explored here. 
In summary, to a large extent our whole approach can be formulated in a language with two light flavors. 
Conceptually this connects 
our present work to the electromagnetic form factors of the charmed $\Lambda_c$, $\Sigma_c$, $\Sigma^*_c$ 
states\footnote{We thank M.F.M.\ Lutz for pointing out this interesting cross-relation.} \cite{Agashe:2014kda}. 
We do not dwell on this connection any further in the present work.

Let us now summarize our results: 
In the electric sector we have found that the inclusion of the decuplet $\Sigma^*$ resonances is crucial. 
This leads to a large cancelation between $\Sigma$ and $\Sigma^*$ exchange such that even the 
unsubtracted dispersion relation for the electric transition form factor is reasonably satisfied. 
The cancelation leads to a small electric transition radius, 
in qualitative agreement with the direct three-flavor $\chi$PT calculation \cite{Kubis:2000aa}. 
Even quantitative agreement can be 
achieved by a fine-tuning of the coupling constant that governs the decay width of the $\Sigma^*$ resonances.

The inclusion of the decuplet for the calculation of the electric radii has also been discussed in \cite{Puglia:2000jy} 
in the heavy-baryon limit of $\chi$PT. There the decuplet did not contribute to the $\Sigma$-$\Lambda$ transition radius. 
Obviously this is different in our relativistic set-up. 

As a further consequence of the cancelation between $\Sigma$ and $\Sigma^*$ exchange 
our electric transition form factor remains small throughout the whole 
low-energy region. This does not fully agree with 
the results of \cite{Kubis:2000aa} obtained in a framework where three-flavor $\chi$PT is augmented with explicit 
vector-meson fields in the antisymmetric tensor representation. We would like to stress again that the inclusion of vector mesons 
in $\chi$PT is not an entirely model-independent procedure --- one of the reasons why we replace this part of the 
framework by dispersion theory. Clearly it remains to be seen how the electric transition form factor
is modified, if the hadronic input is pushed to higher orders and/or the kaon inelasticity is included. 

The magnetic sector is sensitive to a low-energy constant of NLO $\chi$PT. Without the inclusion of such a term there remains 
an ambiguity how to properly treat the spin-3/2 $\Sigma^*$ resonances in a relativistic framework (in the electric sector this 
ambiguity is relegated to NNLO). With a reasonable size estimate for this low-energy constant we can reproduce the 
three-flavor $\chi$PT prediction \cite{Kubis:2000aa} for the magnetic transition radius. In turn by a measurement of the 
magnetic transition radius this low-energy constant could be narrowed down, leading to a significant increase of predictive
power for the low-energy shape of the magnetic transition form factor. Such a measurement could be feasible at FAIR by 
determining the differential decay width of $\Sigma^0 \to \Lambda \, e^+ e^-$. In general, this Dalitz decay depends on two 
variables, for instance the dilepton mass and the angle between $\Lambda$ and electron in the dilepton rest 
frame \cite{husek-leupold}. However, if the electric transition radius is small --- as suggested by our results and 
the ones from \cite{Kubis:2000aa} ---, then a decay width that is just one-fold differential in the dilepton invariant mass 
would be sufficient to extract the magnetic transition radius. A back-to-the-envelope estimate shows that the effort to extract 
a magnetic transition radius from $\Sigma^0 \to \Lambda \, e^+ e^-$ is comparable to the extraction of the slope of the 
pion transition form factor from $\pi^0 \to \gamma \, e^+ e^-$; see \cite{Adlarson:2016ykr} for a recent experiment and 
\cite{Hoferichter:2014vra} for a recent dispersive calculation. 

The framework presented here can be extended to all isovector form factors and transition form factors of the spin-1/2 and 
spin-3/2 baryon ground states. For the octet and decuplet states one might use the very same three-flavor chiral input for the 
baryonic scattering amplitudes. This might provide some additional cross-relations between observables that we have not 
utilized so far with our focus on just the one transition from $\Sigma^0$ to $\Lambda$. 
On the other hand, if one wants to be more 
conservative, one might just use isospin symmetry and a two-flavor chiral input separately for each strangeness sector. 
The complementary experimental program would be a dedicated study of the differential distributions of 
hyperon Dalitz decays $Y \to Y' \, e^+ e^-$. With data input and dispersion theory the experimentally hardly accessible 
space-like region can be addressed, which in turn could provide a new angle on the structure of baryons.

\begin{acknowledgement}
We thank K.\ Sch\"onning for initiating this work by her questions about hyperon form factors. We also thank O.\ Junker 
for cross-checking some of our results, B.\ Kubis and P.\ Salabura for valuable discussions and encouragement 
and J.R.\ Pelaez, J.\ Ruiz de Elvira and G.\ Colangelo, P.\ Stoffer for providing us with their respective pion phase shifts. \\
While finishing this work we became aware of the conference proceedings \cite{Alarcon:2017lkk}. Here the peripheral structure 
of hyperons is addressed by combining dispersion theory with leading-order chiral perturbation theory. It will be interesting 
to compare our results once the details of \cite{Alarcon:2017lkk} are published.
\end{acknowledgement}

%\newpage

\appendix

\section{Cross-check with $\Delta$ exchange and large-$N_c$ relations}
\label{sec:Delta}

As a cross-check of the results (\ref{eq:bornfeyn}), (\ref{eq:resfeyn}) we have also calculated the 
nucleon- and Delta-exchange contributions to $p \bar n \to \pi^+ \pi^0$. The nucleon exchange is obtained from 
(\ref{eq:bornfeyn}) by replacing all the hyperon masses by the nucleon mass $m_N$ and by 
\begin{eqnarray}
  \label{eq:replbornDelta}
  \frac{D F}{\sqrt{3}} \to - \frac{\sqrt{2} \, g_A^2}{4}  \,.
\end{eqnarray}
The Delta exchange is obtained from (\ref{eq:resfeyn}) by replacing the hyperon masses by $m_N$, the $\Sigma^*$ mass by 
$m_\Delta$ and 
\begin{eqnarray}
  \label{eq:replbornDelta2}
  -\frac{h_A^2}{\sqrt{3}} \to \frac{2 \sqrt{2} \, h_A^2}{3}  \,.
\end{eqnarray}
In the large-$N_c$ limit the leading-order contributions to the electric amplitude from nucleon exchange and Delta exchange 
should cancel
each other. The nucleon-exchange contribution to the magnetic amplitude should be twice the one from Delta exchange 
(here with same sign); see, e.g., the corresponding discussion in \cite{Granados:2013moa} and references therein. 

This is indeed what one finds. We first note that the nucleon exchange does not contain the structure $\bar v_n u_p$. Therefore 
one obtains $T_M^N = T_E^N$. For both the nucleon and the Delta exchange the pole terms dominate the polynomial terms in the 
large-$N_c$ limit. This is satisfying as the polynomial terms for the Delta exchange depend on the spurious spin-$1/2$ modes.
In the large-$N_c$ limit the masses of nucleon and Delta are degenerate. 
Keeping only the pole terms yields
\begin{eqnarray}
  {\cal M}_N  & = & - \frac{\sqrt{2} \, g_A^2}{2 F_\pi^2} \, 
    \bar v_n \gamma^\mu k_\mu u_p \, m_N^2 \, \nonumber \\ && \times 
    \left( \frac{1}{t-m_N^2}+\frac{1}{u-m_N^2} \right) 
  \label{eq:bornfeynln}
\end{eqnarray}
and
\begin{eqnarray}
  && {\cal M}_\Delta  = -\frac{\sqrt{2} \, h_A^2}{4 \cdot 9 \, F_\pi^2} \, \nonumber \\ && \times \Bigg(
    \bar v_n u_p \, 3 \, (s-2m_\pi^2) \, m_N \,
    \left( \frac{1}{t-m_N^2}-\frac{1}{u-m_N^2} \right) \nonumber \\ && \phantom{m}{}
    + \bar v_n \gamma^\mu k_\mu u_p \, 2 \, m_N^2 \,
    \left( \frac{1}{t-m_N^2}+\frac{1}{u-m_N^2} \right) \Bigg)  \,.
  \label{eq:resfeynln}
\end{eqnarray}
Recalling the large-$N_c$ relation $h_A=3 g_A/\sqrt{2}$ and the fact that the $\bar v_n u_p$ term does not contribute to the 
magnetic amplitude, we observe that indeed $T_M^\Delta = T_M^N/2$. 

The calculation of the electric amplitude is slightly more complicated. We will show in the following that for the Delta 
exchange the $\bar v_n u_p$ term contributes $(-3)$ times the other term. In this way the Delta contribution to the electric 
amplitude is given by $-3+1 = -2$ times the $\bar v_n \gamma^\mu k_\mu u_p$ structure. This in turn leads to 
$T_E^\Delta = -T_E^N$ as it should be. 

We address the ratio of the two terms in (\ref{eq:resfeynln}) for the electric case. We obtain
\begin{eqnarray}
  && \frac{\bar v_n u_p \, 3 \, (s-2m_\pi^2) \, m_N \, (u-t)}{\bar v_n \gamma^3 k_3 u_p \, 2 \, m_N^2 \, (u+t-2m_N^2)}
  \nonumber \\
  && \approx  \frac{m_N \, 3 \, (s-2m_\pi^2) \, m_N \, (u-t)}{p_z \, (-2 p_{\rm c.m.} \, \cos\theta) \, 2 \, m_N^2 \, (u+t-2m_N^2)} 
  \nonumber \\
  && =  \frac{3 \, (s-2m_\pi^2)}{u+t-2m_N^2} = -3  
  \label{eq:longcalc1}
\end{eqnarray}
as claimed.

\section{Cut structure of resonance pole terms}
\label{sec:app-cutres}

The $t$- and $u$-channel pole terms cause left-hand cuts in the $\bar \Lambda \, \Sigma \to 2 \pi$ amplitudes. 
It is worth to figure out in particular the cut structure of the resonance-exchange terms since a meaningful matching 
of these terms to $\chi$PT requires to avoid as much as possible the cuts and the effects/structures caused by them. 
The starting points of the cuts are given by solutions of the equation
\begin{equation}
  \label{eq:ABcuts}
  \left( -\frac12 \, s + \frac12 \, m_\Sigma^2 + \frac12 \, m_\Lambda^2 + m_\pi^2 -m_{\Sigma^*}^2 \right)^2 
  - 4 \, p_z^2 \, p_{\rm c.m.}^2 = 0 \,.
\end{equation}
Analytical formulae could be provided for these starting points of the cuts (solutions of a quadratic equation in $s$), but 
we will not present them here. Instead we present the numerical values for points that are of relevance for the 
resonance-exchange contributions to the $\bar \Lambda \, \Sigma \to 2 \pi$ amplitudes. The (pseudo-)thresholds
are given by 
\begin{eqnarray}
  \label{eq:s123}
    && s_1=(m_\Sigma-m_\Lambda)^2 \approx 0.00592\,\mbox{GeV}^2  \,, \nonumber \\ 
    && s_2=(2 m_\pi)^2 \approx 0.0779\,\mbox{GeV}^2 \,,  \nonumber \\
    && s_3=(m_\Sigma+m_\Lambda)^2 \approx 5.33\,\mbox{GeV}^2 \,. 
\end{eqnarray}
There are two cuts. One ranges from $-\infty$ to $s_{c1} \approx -0.105$ GeV$^2$, 
the other from $s_{c2} \approx -0.00305\,$GeV$^2$ to zero. The magnetic amplitude has cusps at $s_{c1}$, $s_{c2}$ and zero. 
(The electric amplitude diverges at $s_{c1}$ and $s_{c2}$.) 
The very small cut from 
$s_{c2}$ to zero has (only) an impact on the region around $s=0$. Thus a matching to $\chi$PT should be performed for small 
values of $s$ but for $\vert s \vert \gg \vert s_{c2} \vert$. One possibility is a matching at the 
two-pion threshold $s_2=(2 m_\pi)^2$. A second possible choice is a matching for negative $s$ with $s_{c1} \ll s \ll s_{c2}$. 
For the latter case a reasonable choice might be to use the threshold value for 
$t$ of the reaction $\Sigma \, \pi \to \Lambda \, \pi$. The threshold of this reaction is at $m_\Sigma+m_\pi$ and there 
the value of the momentum transfer from the pions to the baryons is given by 
\begin{equation}
  \label{eq:st0}
  s_{t0}=-m_\pi \, (m_\Sigma^2-m_\Lambda^2)/(m_\Sigma+m_\pi) \approx -0.0186\,\mbox{GeV}^2 \,,
\end{equation}
which lies comfortably between $s_{c1}$ and $s_{c2}$. 

A third possibility is a theoretical calculation with an appropriate 
low-energy limit. In the limit $m_\pi \to 0$ the small cut between $s_{c2}$ and zero disappears. Only one cut is left that 
ranges from $-\infty$ to 
\begin{eqnarray}
  \label{eq:cut2flchirallim}
  s_c = -\frac{m_{\Sigma^*}^4-(m_\Sigma^2 + m_\Lambda^2)\, m_{\Sigma^*}^2 + m_\Sigma^2 \, m_\Lambda^2}{m_{\Sigma^*}^2}  \,.
\end{eqnarray}
Thus in this limit one can perform a matching at $s=0$. 
If one puts $m_\pi$ to zero it makes sense to also neglect the mass difference between $m_\Sigma$ and $m_\Lambda$. Otherwise the 
two pseudo-thresholds at $s_1$ and $s_2$ would change their ordering. After this double limit $m_\pi \to 0$ and 
$m_\Lambda \to m_\Sigma$ one can evaluate the resonance-pole contribution to the magnetic amplitude for $s=0$. One obtains 
\begin{eqnarray}
  && K^M_{\rm res,low} := \lim_{s \to 0} \, \lim_{m_\Lambda \to m_\Sigma} \, \lim_{m_\pi \to 0} K^M_{\rm res}(s) \nonumber \\ &&
  = \frac{h_A^2}{24 \sqrt{3} F_\pi^2} \, 
  \frac{(-m_{\Sigma^*}^2+4 m_{\Sigma^*} m_\Sigma - m_\Sigma^2) \, (m_{\Sigma^*} + m_\Sigma)}{m_{\Sigma^*}^2 \, (m_{\Sigma^*} - m_\Sigma)} \,.
  \nonumber \\ 
  \label{eq:kmagreslowdef}
\end{eqnarray}
Matching is then performed according to (\ref{eq:fixpolyEM}).

A detailed inspection of the magnetic amplitude (not displayed here) shows that the ``distortion'' of the curve 
caused by the small cut is 
of minor importance at $s_{t0}$ and at $s_2$. In the vicinity of $s_1$ there are no cuts, but there is already a large slope. 
Thus a matching to $\chi$PT at $s_1$ would not be a good choice. Numerically the ratio between the amplitudes at 
$s_2$ and at $s_{t0}$ is about 
0.78. The ratio between the magnetic amplitude at $s_{t0}$ and (\ref{eq:kmagreslowdef}) is about 0.97, i.e.\ very close to 1, 
demonstrating agreement between the idea to match at the physically reasonable point $s_{t0}$ and the theoretical low-energy 
calculation (\ref{eq:kmagreslowdef}). For our numerical results we will use the analytical expression (\ref{eq:kmagreslowdef}).
Note, however, that in view of the large uncertainties in $b_{10}$ a matching at the two-pion threshold would also be a 
reasonable choice.

\bibliography{lit}{}
\bibliographystyle{epj}
\end{document}